\documentclass[12pt,preprint]{aastex}
\def\bm#1{\mbox{\boldmath $#1$}} %--- vector (bold face)
\usepackage{pslatex}
\usepackage{graphicx}

\begin{document}

\title{Strong Anisotropic MHD Turbulence with Cross Helicity}
\author{Benjamin D. G. Chandran}
\email{benjamin.chandran@unh.edu} 
\affil{Space Science Center and
Department of Physics, University of New Hampshire}

\begin{abstract}
This paper proposes a new phenomenology for strong incompressible MHD
turbulence with nonzero cross helicity. This phenomenology is then
developed into a quantitative Fokker-Planck model that describes the
time evolution of the anisotropic power spectra of the fluctuations
propagating parallel and anti-parallel to the background magnetic
field~$\bm{B}_0$.  It is found that in steady state the power spectra
of the magnetic field and total energy are steeper than
$k_\perp^{-5/3}$ and become increasingly steep as $C/{\cal E}$
increases, where $C=\displaystyle \int d^3\! x \; \bm{v}\cdot\bm{B}$
is the cross helicity, ${\cal E}$ is the fluctuation energy, and
$k_\perp$ is the wavevector component perpendicular to~$\bm{B}_0$.
Increasing $C$ with fixed ${\cal E}$ increases the time required for
energy to cascade to smaller scales, reduces the cascade power,
and increases the anisotropy of the small-scale
fluctuations. The implications of these results for the solar wind and
solar corona are discussed in some detail.
\end{abstract}

\maketitle

\section{Introduction}

Much of our current understanding of incompressible
magnetohydrodynamic (MHD) turbulence has its roots in the pioneering
work of Iroshnikov~(1963) and Kraichnan~(1965). These studies
emphasized the important fact that Alfv\'en waves travelling in the
same direction along a background magnetic field do not interact with
one another and explained how one can think of the cascade of energy
to small scales as resulting from collisions between oppositely
directed Alfv\'en wave packets.  They also argued that in the absence
of a mean magnetic field, the magnetic field of the energy-containing
eddies at scale~$L$ affects fluctuations on scales~$\ll L$ much in the
same way as would a truly uniform mean magnetic field.

Another foundation of our current understanding is the finding that
MHD turbulence is inherently anisotropic.  Montgomery \& Turner~(1981)
and Shebalin, Matthaeus, \& Montgomery~(1983) showed that a strong
uniform mean magnetic field~$\bm{B}_0$ inhibits the cascade of energy
to small scales measured in the direction parallel to~$\bm{B}_0$. This
early work was substantially elaborated upon by Higdon~(1984),
Goldreich \& Sridhar~(1995, 1997), Montgomery \& Matthaeus~(1995), Ng
\& Bhattacharjee~(1996, 1997), Galtier et al~(2000), Cho \&
Lazarian~(2003), Oughton et al~(2006), and many others. For example,
Cho \& Vishniac (2000) used numerical simulations to show that when
the fluctuating magnetic field~$\delta B$ is $\ga B_0$ the small-scale
turbulent eddies become elongated along the local magnetic field
direction.  Goldreich \& Sridhar (1995) introduced the important and
influential idea of ``critical balance,'' which holds that at each
scale the linear wave period for the bulk of the fluctuation energy is
comparable to the time for the fluctuation energy to cascade to
smaller scales. Goldreich \& Sridhar (1995, 1997), Maron \&
Goldreich~(2001), and Lithwick \& Goldreich~(2001) clarified a number
of important physical processes in anisotropic MHD turbulence and used
the concept of critical balance to determine the ratio of the
dimensions of turbulent eddies in the directions parallel and
perpendicular to the local magnetic field.

Over the last several years, research on MHD turbulence has been
proceeding along several different lines. For example, one group of
studies has attempted to determine the power spectrum, intermittency,
and anisotropy of strong incompressible MHD turbulence using direct
numerical simulations. (See, e.g., Cho \& Vishniac 2000, M\"uller \&
Biskamp 2000, Maron \& Goldreich 2001, Cho et~al~2002, Haugen
et~al~2004, Muller \& Grappin~2005, Mininni \& Pouquet 2007, Perez \&
Boldyrev 2008).  Another series of papers has explored the properties
of anisotropic turbulence in weakly collisional magnetized plasmas
using gyrokinetics, a low-frequency expansion of the Vlasov equation
that averages over the gyromotion of the particles.  (Howes et
al~2006, 2007a, 2007b; Schekochihin et al 2007). These authors
investigated the transition between the Alfv\'en-wave cascade and a
kinetic-Alfv\'en-wave cascade at length scales of order the proton
gyroradius~$\rho_i$, as well as the physics of energy dissipation and
entropy production in the low-collisionality regime. Turbulence at
scales~$\lesssim \rho_i$ has also been examined both numerically and
analytically within the framework of fluid models, in particular Hall
MHD and electron MHD.  (Biskamp, Schwarz, \& Drake 1996, Biskamp
et~al~1999, Matthaeus et al~2003; Galtier \& Bhattacharjee 2003, 2005;
Cho \& Lazarian 2004; Brodin et al~2006, Shukla et al~2006).  Another
group of studies has investigated the power spectrum, intermittency,
and decay time of compressible MHD turbulence.  (Oughton et al 1995,
Stone et~al~1998, Lithwick \& Goldreich~2001, Boldyrev et al~2002,
Padoan et al~2004, Elmegreen \& Scalo 2004).  Additional work by
Kuznetsov (2001), Cho \& Lazarian (2002, 2003), Chandran (2005), and
Luo \& Melrose (2006) has begun to address the way in which Alfv\'en
waves, fast magnetosonic waves, and slow magnetosonic waves interact
in compressible weak MHD turbulence. Another recent development is the
finding that strong incompressible MHD turbulence leads to alternating
patches of alignment and anti-alignment between the velocity and
magnetic-field fluctuations. (Boldyrev 2005, 2006; Beresnyak \&
Lazarian 2006; Mason, Cattaneo, \& Boldyrev 2006; Matthaeus
et~al~2007) These studies examined how the degree of local alignment
(and anti-alignment) depends upon length scale, as well as the effects of
alignment upon the energy cascade time and the power spectrum of the
turbulence.

The topic addressed in this paper is the role of cross helicity
in incompressible MHD turbulence. The cross helicity is defined as
\begin{equation}
C = \int d^3\!x \; \bm{v}\cdot {B},
\label{eq:defC} 
\end{equation} 
where $\bm{v}$ is the velocity and~$\bm{B}$ is the magnetic field. The
cross helicity is conserved in the absence of dissipation and can be
thought of as a measure of the linkages between lines of vorticity and
magnetic field lines, both of which are frozen to the fluid flow in
the absence of dissipation (Moffatt 1978).  In the presence of a
background magnetic field, $\bm{B}_0 = B_0 \hat{z}$, the cross
helicity is also a measure of the difference between the energy of
fluctuations travelling in the $-z$ and $+ z$ directions.  Dobrowolny,
Mangeney, \& Veltri~(1980) showed that MHD turbulence with cross
helicity decays to a maximally aligned state, with $\delta \bm{v} =
\pm \delta\bm{ B}/\sqrt{4\pi\rho}$, where $\delta \bm{v}$ and $\delta
\bm{B}$ are the fluctuating velocity and magnetic field and $\rho$ is
the mass density. Different decay rates for the energy and cross
helicity were also demonstrated by Matthaeus \& Montgomery~(1980).  In
another early study, Grappin, Pouquet, \& L\'eorat~(1983) used a
statistical closure, the eddy-damped quasi-normal Markovian (EDQNM)
approximation, to study strong 3D incompressible MHD turbulence with
cross helicity, assuming isotropic power spectra.  They found that
when~$C\neq 0$, the total energy spectrum is steeper than the
isotropic Iroshnikov-Kraichnan $k^{-3/2}$~spectrum. Pouquet
et~al~(1988) found a similar result in direct numerical simulations of
2D incompressible MHD turbulence.  More recently, Lithwick, Goldreich,
\& Sridhar~(2007) and Beresnyak \& Lazarian (2007) addressed the role
of cross helicity in strong MHD turbulence taking into account the
effects of anisotropy.

This paper presents a new phenomenology for strong, anisotropic,
incompressible MHD turbulence with nonzero cross helicity, and is
organized as follows. Section~\ref{sec:wpc} presents some relevant
theoretical background. Section~\ref{sec:theory} introduces the new
phenomenology as well as two nonlinear advection-diffusion equations
that model the time evolution of the power spectra.  Analytic and
numerical solutions to this equation in the weak-turbulence and
strong-turbulence regimes are presented in Sections~\ref{sec:weak}
and~\ref{sec:strong}.  Section~\ref{sec:strong} also presents a simple
phenomenological derivation of the power spectra and anisotropy in
strong MHD turbulence.  Section~\ref{sec:trans} presents a numerical
solution to the advection-diffusion equation that shows the smooth
transition between the weak and strong turbulence
regimes. Section~\ref{sec:unequal} addresses the case in which the
parallel correlation lengths of waves propagating in opposite
directions along the background magnetic field are unequal at the
outer scale.  In Section~\ref{sec:solarwind}, the proposed
phenomenology is applied to turbulence in the solar wind and solar
corona, and in Section~\ref{sec:comp} the results of this work are
compared to the recent studies of Lithwick, Goldreich, \&
Sridhar~(2007) and Beresnyak \& Lazarian~(2007).

\section{Energy Cascade from Wave-Packet Collisions}
\label{sec:wpc} 

The equations of ideal incompressible MHD can be written
\begin{equation} 
\frac{\partial \bm{w}^\pm}{\partial t} +\left(
\bm{w}^\mp \mp v_{\rm A}\hat{z}  \right) \cdot \nabla \bm{w}^\pm
= -\nabla \Pi 
\label{eq:mhd} 
\end{equation}
where $\bm{w}^\pm = \bm{v} \pm (\delta\bm{B}/\sqrt{4\pi \rho})$ are
the Elsasser variables, $\bm{v}$ is the fluid velocity, $\delta
\bm{B}$ is the magnetic field fluctuation, $\rho$ is the mass density,
which is taken to be uniform and constant, $v_A = B_0/\sqrt{4\pi
  \rho}$ is the Alfv\'en speed, $\bm{B}_0 = B_0 \hat{z}$ is the mean
magnetic field, and $\Pi = (p + B^2/8\pi)/\rho$, which is determined
by the incompressibility condition, $\nabla \cdot \bm{w}^\pm = 0$.
Throughout this paper it is assumed that $\delta B \lesssim B_0$ and
$w^\pm \lesssim v_A$.

In the limit of small-amplitude fluctuations ($w^\pm \ll v_A$), the
nonlinear term $\bm{w}^\mp \cdot \nabla \bm{w}^\pm$ in
equation~(\ref{eq:mhd}) can be neglected to a first approximation, and
the curl of equation~(\ref{eq:mhd}) becomes
\begin{equation}
\left(\frac{\partial }{\partial t} \mp v_A \frac{\partial}{\partial
  z}\right) \nabla \times \bm{w}^\pm = 0,
\end{equation} 
which is solved by setting $\nabla \times \bm{w}^\pm$ equal to an
arbitrary function of $z\pm v_A t$. Thus, $\bm{w}^\pm$ represents
fluctuations with $\bm{v} = \pm \bm{b}$ that propagate in $\mp z$
direction at speed~$v_A$ in the absence of nonlinear interactions.  In
the absence of an average velocity, the cross helicity defined in
equation~(\ref{eq:defC}) can be rewritten as
\begin{equation}
C = \frac{\sqrt{\pi\rho}}{2} \int d^3\! x \, \left[(w^+)^2 - (w^-)^2\right].
\label{eq:defC2} 
\end{equation} 
The cross helicity is thus proportional to the difference in energy
between fluctuations propagating in the $-z$ and $+z$ directions.

\begin{figure}[h]
\includegraphics[width=3in]{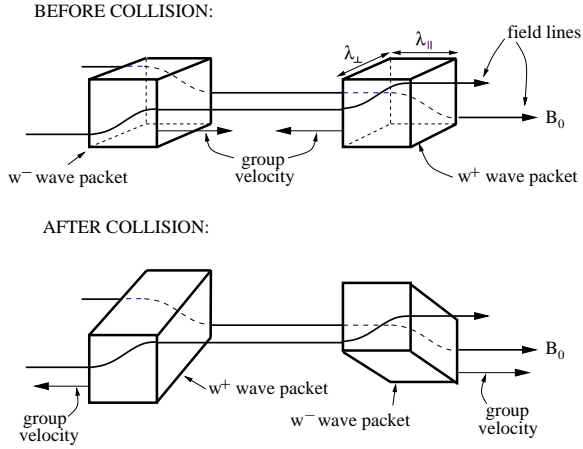}
\caption{\footnotesize  When two wave packets collide, each wave packet
follows the field lines of the other wave packet and gets sheared.
\label{fig:f1}}
\end{figure}

Equation~(\ref{eq:mhd}) shows that the nonlinear term is nonzero only
at those locations where both $w^+$ and $w^-$ are nonzero. Nonlinear
interactions can thus be thought of as collisions between oppositely
directed wave packets (Kraichnan 1965).  When both $w^+$ and $w^-$ are
nonzero, equation~(\ref{eq:mhd}) indicates that the $\bm{w}^\pm$
fluctuations are advected not at the uniform velocity $\mp v_A
\hat{z}$, but rather at the non-uniform velocity $\mp v_A\hat{z} +
\bm{w}^\mp$.  Maron \& Goldreich (2001) elaborated upon this idea by
showing that to lowest order in fluctuation amplitude, if one neglects
the pressure term, then $w^+$ wave packets are advected along the
hypothetical magnetic field lines corresponding to the sum of
$\bm{B}_0$ and the part of $\delta \bm{B}$ arising from the $w^-$
fluctuations.  This result can be used to construct a geometrical
picture for how wave-packet collisions cause energy to cascade to
smaller scales, as depicted in Figure~\ref{fig:f1}.  In this figure,
two oppositely directed wave packets of dimension $\sim \lambda_\perp$
in the plane perpendicular to~$\bm{B}_0$ and
length~$\lambda_\parallel$ along~$\bm{B}_0$ pass through one another
and get sheared.  Collisions between wavepackets of
similar~$\lambda_\perp$ are usually the dominant mechanism for
transferring energy from large scales to small scales.  The duration
of the collision illustrated in the figure is approximately~$\Delta t
\sim \lambda_\parallel/v_A$.  The fluctuating velocity and
magnetic field are taken to be in the plane perpendicular
to~$\bm{B}_0$, as is the case for linear
shear Alfv\'en waves.  The magnitude of the nonlinear term in
equation~(\ref{eq:mhd}) is then $\sim w^+_{\lambda_\perp}
w^-_{\lambda_\perp}/\lambda_\perp$, where $w^\pm_{\lambda_\perp}$ is
the rms amplitude of the $w^\pm$ wave packet. The fractional change in
the $\bm{v}$ and $\bm{b}$ fields of the $w^\mp$ wave packet induced by
the collision is then roughly
\begin{equation}
\left(\frac{w^+_{\lambda_\perp} w^-_{\lambda_\perp}}{\lambda_\perp}\right)\times 
\left(\frac{\Delta t}{
w^\mp_{\lambda_\perp}}\right) = \frac{w_{\lambda_\perp}^\pm \lambda_\parallel}{v_A \lambda_\perp}.
\label{eq:defchi1} 
\end{equation} 

If this fractional change is $\ll 1$ for both $w^+$ and $w^-$
fluctuations then neither wave packet is altered significantly by a
single collision, and the turbulence is weak. Wave packets travel a
distance~$\gg \lambda_\parallel$ before being significantly distorted,
and the fluctuations can thus be viewed as linear waves that are only
weakly perturbed by nonlinear interactions with other waves.  In the
wave-packet collision depicted in Figure~\ref{fig:f1}, the right-hand
side of the $ w^-$ wave packet is altered by the collision in almost
the same way as the left-hand side, since both sides encounter
essentially the same~$w^+$ wave packet, since the~$w^+$ packet is
changed only slightly during the collision.  Changes to the profile of
a wave packet along the magnetic field are thus weaker than changes in
the profile of a wave packet in the plane perpendicular to~$\bm{B}_0$
(Shebalin et al 1983, Ng \& Bhattacharjee 1997, Goldreich \& Sridhar
1997, Bhattacharjee \& Ng~2001, Perez \& Boldyrev~2008). As a result,
in the weak-turbulence limit, the cascade of energy to
small~$\lambda_\parallel$ is much less efficient than the cascade of
energy to small~$\lambda_\perp$ (Galtier et al 2000).

On the other hand, if the fractional change in
equation~(\ref{eq:defchi1}) is of order unity then a $w^\mp$ wave
packet is distorted substantially during a single collision, and the
turbulence is said to be ``strong.'' In the case that the fractional
change in equation~(\ref{eq:defchi1}) is $\sim 1$ for one fluctuation
type, (e.g., $w^-$) but $\ll 1$ for the other ($w^+$), the turbulence
is still referred to as strong. It should be noted that strong
turbulence can arise when $w^\pm_{\lambda_\perp} \ll v_A$, provided
that $\lambda_\perp \ll \lambda_\parallel$.  In strong turbulence
energy cascades to smaller~$\lambda_\parallel$ to a greater extent
than in weak turbulence, but the primary direction of energy flow
in~$k$-space is still to larger~$k_\perp$, as discussed in the next
section.

\section{Anisotropic MHD Turbulence with Cross Helicity}
\label{sec:theory} 

In order to develop an analytical model, it is convenient to work 
in terms of the Fourier transforms of the fluctuating $w^\pm$ fields, given by
\begin{equation}
\tilde{\bm{w}}^\pm(\bm{k}) = \frac{1}{(2\pi)^3}\int d^3\!x \; \bm{w}(\bm{x})
e^{-i\bm{k}\cdot \bm{x}}.
\end{equation} 
The three-dimensional power spectrum~$A^\pm(\bm{k})$ is defined by the equation
\begin{equation}
\langle \tilde{\bm{w}}^\pm (\bm{k}) \cdot \tilde{\bm{w}}^\pm(\bm{k}_1)\rangle
= A^\pm(\bm{k}) \delta(\bm{k}+ \bm{k}_1),
\end{equation} 
where $\langle \dots \rangle$ denotes an ensemble average.
Cylindrical symmetry about~$\bm{B}_0$ is assumed, so that
$A^\pm(\bm{k}) = A^\pm (k_\perp,k_\parallel)$, where
$k_\perp$ and $k_\parallel$ are the components of $\bm{k}$
perpendicular and parallel to~$\bm{B}_0$.
The mean-square velocity associated with $w^\pm$ fluctuations
is then 
\begin{equation}
(\delta v^\pm)^2 = \frac{1}{4}\int d^3 \!k \; A^\pm(k_\perp,k_\parallel).
\label{eq:defdeltav} 
\end{equation} 

It is assumed
that at each value of~$k_\perp$ there is a parallel wave number
$\overline{k}_\parallel ^{\,\pm}(k_\perp)$ such that
(1) the bulk of the $w^\pm$  fluctuation energy is 
at $|k_\parallel| < \overline{k}_\parallel ^{\,\pm}(k_\perp)$
and (2) $A^\pm(k_\perp,k_\parallel)$ depends only weakly 
on~$k_\parallel$ for $|k_\parallel| < \overline{k}_\parallel ^{\,\pm}(k_\perp)$.
A $w^\pm$
wavepacket at perpendicular scale $k_\perp^{-1}$ then has a correlation
length in the direction of the mean field of $\sim
\left(\overline{k}_\parallel^{\,\pm} \right)^{-1}$.  The rms amplitude
of the fluctuating Elsasser fields at a perpendicular scale
$k_\perp^{-1}$, denoted $w_{k_\perp}^\pm$, is given by
\begin{equation}
(w^\pm_{k_\perp})^2 \sim A^\pm(k_\perp,0)k_\perp^2 \overline{k}_\parallel^{\,\pm}.
\label{eq:defdeltakpar} 
\end{equation} 
As described in the section~\ref{sec:wpc}, when a $w^\mp$ wave packet
at scale $k_\perp^{-1}$ collides with a $w^\pm$ wave packet at scale
$k_\perp^{-1}$, the fractional change in the $w^\mp$ packet resulting
from the collision is approximately
\begin{equation}
\chi_{k_\perp}^\pm = \frac{k_\perp w_{k_\perp}^\pm}{{\overline k}_\parallel ^{\,\pm} v_A}.
\label{eq:defchi2} 
\end{equation} 
The wave number $k_c^\pm$ is defined to be the value of $\overline{k}_\parallel^{\,\pm}$
for which $\chi_{k_\perp}^\pm = 1$. Thus,
\begin{equation}
k_c^\pm = \frac{k_\perp^4 A^\pm(k_\perp, 0)}{v_A^2}.
\label{eq:defkc} 
\end{equation} 

\subsection{The energy cascade time}

When $\overline{k}_\parallel^{\,-} \gg k_c^-$, the value of
$\chi_{k_\perp} ^-$ is~$\ll 1$ and a $w^+$ is only weakly affected by
a single collision with a $w^-$ wave packet. Each such collision
requires a time $(\overline{k}_\parallel^{\,-} v_A)^{-1}$.  The
effects of successive collisions add incoherently, and thus
$(\chi_{k_\perp}^-)^{-2}$ collisions are required for the $w^+$ wave
packet to be strongly distorted, and for its energy to pass to smaller
scales. The cascade time $\tau_{k_\perp}^+$ for a $w^+$ wave packet at
perpendicular scale~$k_\perp^{-1}$ is
thus roughly
\begin{equation}
\tau_{k_\perp}^+ \sim (\overline{k}_\parallel^{\,-} v_A)^{-1} (\chi_{k_\perp}^-)^{-2}
\sim \frac{1}{k_c^- v_A} \mbox{ \hspace{0.3cm} (weak turbulence).}
\label{eq:tau1} 
\end{equation} 
Similarly, if $\chi_{k_\perp}^+ \ll 1$, then $\tau_{k_\perp}^- \sim (k_c^+ v_A)^{-1}$.

When $\overline{k}_\parallel^{\,-} \sim k_c^-$, the value of $\chi_{k_\perp}
^-$ is~$\sim 1$, a $w^+$ is strongly distorted during a single
wave packet collision, and the turbulence is strong. Each such
collision takes a time $(\overline{k}_\parallel^{\,-}
v_A)^{-1}$. Since $\overline{ k}_\parallel^{\,-} \sim k_c^-$, 
\begin{equation}
\tau_{k_\perp}^+ \sim (\overline{k}_\parallel^{\,-}
v_A)^{-1} \sim \frac{1}{k_c^- v_A} \mbox{ \hspace{0.3cm} (strong turbulence).} 
\label{eq:tau2} 
\end{equation} 
Similarly, if $\chi_{k_\perp}^+ \sim 1$, then $\tau_{k_\perp}^- \sim
(k_c^+ v_A)^{-1}$.

The case $\overline{k}_\parallel^{\,\pm} \ll k_c^\pm$ (i.e.,
$\chi_{k_\perp}^\pm \gg 1$) is explicitly excluded from the
discussion.  Initial conditions could in principle be set up in which
$\overline{k}_\parallel^{\,\pm} \ll k_c^\pm$. However, the cascade
mechanisms described in section~\ref{sec:parcasc} will not produce the
condition $\overline{k}_\parallel^{\,\pm} \ll k_c^\pm$ if it is not
initially present.  It should be emphasized that in both weak
turbulence and strong turbulence, the cascade time is given by the
same formula, $\tau_{k_\perp}^{\pm} \sim (k_c^\mp v_A)^{-1}$, which
involves the $A^\mp$ spectrum evaluated only at $k_\parallel=0$. 

\subsection{The Cascade of Energy to Larger~$k_\parallel$}
\label{sec:parcasc} 

The two basic mechanisms for transferring fluctuation energy to
larger~$k_\parallel$ were identified by Lithwick, Goldreich, \&
Sridhar~(2007).  The
first of these can be called ``propagation with distortion.''  Suppose
a $w^+$ wave packet of perpendicular scale $k_\perp^{-1}$ and
arbitrarily large initial parallel correlation length begins colliding
at $t=0$ with a stream of $w^-$ wave packets of similar perpendicular
scale.  At time $t=\tau_{k_\perp}^+$, the leading edge of the $w^+$
wave packet has been distorted substantially by the stream of $w^-$
wave packets, but the trailing portion of the $w^+$ wave packet at
distances $\ga 2 v_A \tau_{k_\perp}^+$ behind the leading edge has not
yet encountered the stream of $w^-$ wave packets. If the parallel
correlation length of the~$w^+$ wave packet is initially $ >2 v_A
\tau_{k_\perp}^+$, then during a time~$\tau_{k_\perp}^+$ the $w^+$
wave packet acquires a spatial variation in the direction of the
background magnetic field of length scale~$\sim 2 v_A \tau_{k}^+ \sim
2 (k_c^-)^{-1}$.  This process is modeled as diffusion of $w^\pm$
fluctuation energy in the $k_\parallel$ direction with diffusion
coefficient $D_\parallel^\pm \sim (\Delta k_\parallel)^2/\Delta t$,
where $\Delta k_\parallel = k_c^\mp$ and $\Delta t = \tau_k^\pm$.
``Propagation with distortion'' then leads to a value of
$D_\parallel^\pm$ of $\sim (k_c^\mp)^3 v_A$.

The second mechanism identified by Lithwick, Goldreich, \&
Sridhar~(2007) can be called ``uncorrelated cascade.'' Consider a
$w^+$ wave packet of perpendicular scale~$k_\perp^{-1}$ and
arbitrarily large parallel correlation length, and consider two points
within the wave packet, $P_1$ and $P_2$, that move with the wave
packet at velocity~$-v_A \bm{\hat{z}}$ and are separated by a distance
along~$\bm{B}_0$ of $ 2v_A \tau_{k_\perp}^- \sim 2(k_c^+)^{-1}$.  The
$w^-$ wave packets at perpendicular scale~$k_\perp^{-1}$ encountered
by the portions of the $w^+$ wave packet at $P_1$ and $P_2$ are then
uncorrelated, because $w^-$ wave packets are substantially distorted
while propagating between $P_1$ and $P_2$. Thus, the way in which the
$w^+$ wave packet cascades at location~$P_1$ is not correlated with
the way in which the~$w^+$ wave packet cascades at location~$P_2$. If
the parallel correlation length of the~$w^+$ wave packet is
initially~$>2v_A \tau_{k_\perp}^-$, then wave-packet collisions
introduce a spatial variation along~$\bm{B}_0$ into the $w^+$ wave
packet of length scale~$\sim 2v_A \tau_{k_\perp}^- \sim 2(k_c^+)^{-1}$
during a time~$\tau_{k_\perp}^+$.  Again, we model this as diffusion
of $w^\pm$ fluctuation energy in the $k_\parallel$ direction with
$D_\parallel^\pm \sim (\Delta k_\parallel)^2/\Delta t$ and $\Delta t =
\tau_{k_\perp}^\pm$, but now $\Delta k_\parallel = k_c^\pm$.
``Uncorrelated cascade'' thus leads to a $k_\parallel$-diffusion
coefficient of~$\sim (k_c^\pm)^2 k_c^\mp v_A$.

Accounting for both mechanisms, one can write
\begin{equation}
D^\pm_\parallel \sim (k_{c,\rm max})^2 k_c^\mp v_A,
\label{eq:Dpar} 
\end{equation} 
where $k_{c,\rm max}(k_\perp)$ is the larger of $k_c^+(k_\perp)$ and
$k_c^-(k_\perp)$.  If $k_c^+ > k_c^-$, then $w^+$ energy
diffuses in~$k_\parallel$ primarily through the ``uncorrelated
cascade'' mechanism, while $w^-$ energy
diffuses in~$k_\parallel$ primarily through
the ``propagation with distortion cascade'' mechanism.

\subsection{Advection-Diffusion Model for the Power Spectra}

The phenomenology described in the preceding sections is encapsulated 
by the following nonlinear advection-diffusion equation, 
\begin{equation}
\frac{\partial A^\pm_k}{\partial t} = - \frac{1}{k_\perp}
\frac{\partial }{\partial k_\perp} \left(\frac{c_1 k_\perp^2 A^\pm_k h^\pm_k}{\tau^\pm _{{\rm eff,} k_\perp}}\right) + c_2 (k_{c,\rm max})^2 k_c^\mp v_A \frac{\partial^2 A^\pm_k}{\partial k_\parallel^2} + S^\pm_k - \gamma^\pm_k A^\pm_k,
\label{eq:FPpm} 
\end{equation} 
where $A^\pm_k$ is shorthand for $A^\pm(k_\perp, k_\parallel)$, $c_1$
and $c_2$ are dimensionless constants of order unity, and $S^\pm_k$
and $- \gamma^\pm_k A^\pm_k$ are forcing and damping terms,
respectively.  The first term on the right-hand side of
equation~(\ref{eq:FPpm}) represents advection of fluctuation energy to
larger~$k_\perp$, while the second term represents diffusion of
fluctuation energy to larger~$|k_\parallel|$.  The quantity
$\tau_{{\rm eff}, k_\perp}^\pm$ is an effective cascade time at
perpendicular scale~$k_\perp^{-1}$.  Usually, the transfer of energy
to small scales is dominated by local interactions in $k$-space, and
the cascade time for a $w^+$ wave packet is $\sim (k_c^\mp
v_A)^{-1}$. In some cases, however, the shearing of small-scale wave
packets by much larger-scale wave packets can become important. To
account for such cases, the effective cascade time is taken to be
\begin{equation}
(\tau_{{\rm eff}, k_\perp}^\pm)^{-1}= \max\left[
\frac{q_\perp^4 A^\mp (q_\perp,0)}{v_A}\right]
\mbox{ \hspace{0.3cm}  for $0<q_\perp< k_\perp$,}
\label{eq:taueff} 
\end{equation} 
i.e., $(\tau_{{\rm eff}, k_\perp}^\pm)^{-1}$ is the maximum value
of $k_c^\mp v_A$ for all perpendicular wave numbers between zero
and~$k_\perp$.
The flux of $w^\pm$ energy to larger~$k_\perp$ is 
\begin{equation}
\epsilon^\pm(k_\perp) = 2\pi\int_{-\infty}^\infty dk_\parallel \,
\frac{c_1 k_\perp^2 A^\pm_k h^\pm_k}{\tau^\pm _{{\rm eff,} k_\perp}}.
\label{eq:epsilon} 
\end{equation} 
The term $h^\pm _k$ is given by
\begin{equation}
h^\pm _k = - \frac{1}{A^\pm(k_\perp,0)}\frac{\partial}{\partial k_\perp}
\left[k_\perp A^\pm (k_\perp,0)\right],
\label{eq:defh} 
\end{equation} 
and is included so that $\epsilon^\pm$ increases as the $A^\pm$
spectrum becomes a more steeply declining function of~$k_\perp$, in
accordance with weak turbulence theory (Galtier et al 2000, Lithwick
\& Goldreich~2003).  To match the energy flux in weak turbulence
theory in the limit of zero cross helicity, one must set\footnote{The
  value of~$c_1$ in equation~(\ref{eq:valc1}) is a factor of~2 larger
  than the value that follows from the results of Galtier
  et~al~(2000). It appears that this discrepancy results from the
  omission of a factor of~2 in equation~(54) of Galtier
  et~al~(2000). This can be seen by starting from equation~(46) of
  Galtier et~al~(2000) and using the expression on page~1045 of Leith
  \& Kraichnan~(1972) to simplify polar integrals of the form $\int
  d^2 p\, d^2 q \,\delta (\bm{k} - \bm{p} - \bm{q})F(k,p,q)$ for
  two-dimensional wave vectors $\bm{k}$, $\bm{p}$, and $\bm{q}$, where
  $F$ is a function only of the wave-vector magnitudes and the
  integral is over all values of $\bm{p}$ and $\bm{q}$.}
\begin{equation}
c_1 = -\frac{\pi J}{2},
\label{eq:valc1} 
\end{equation} 
where
\begin{equation}
J = \int_1^\infty dx\int_{-1}^1 dy\, \frac{
2 [(x^2-1)(1-y^2)]^{1/2} (1+xy)^2 [8-(x+y)^3] \,\ln[(x+y)/2]}
{(x^2 - y^2)^4} \simeq -1.87.
\label{eq:defJ}
\end{equation}

For simplicity,
\begin{equation}
c_2 = 1.
\label{eq:valc2} 
\end{equation}

\section{Steady-State Weak Turbulence}
\label{sec:weak} 

This section addresses weak turbulence in which
$\overline{ k}_\parallel^{\,+} \sim \overline{ k}_\parallel^{\,-}$
at the outer scale. The weak-turbulence condition, $\chi_{k_\perp}^\pm \ll 1$,
is equivalent to the condition $k_c^\pm 
\ll \overline{ k}_\parallel^{\,\pm}$. Because $k_\parallel$-diffusion involves
a $\Delta k_\parallel \sim k_c^\pm$ during a time~$\tau_{k_\perp}^\pm$,
the $k_\parallel$-increment
over which energy diffuses 
while cascading to larger~$k_\perp$ is much less than the 
breadth of the spectrum in the $k_\parallel$ direction ($\sim 
\overline{ k}_\parallel^{\,\pm}$),
so the $k_\parallel$-diffusion terms can be ignored to a good approximation.
In this case, equation~(\ref{eq:FPpm}) possesses
a steady-state solution in which $\epsilon^+$ and $\epsilon^-$ are constant,
and in which
\begin{equation}
A^\pm_k = g^\pm(k_\parallel) k_\perp^{-n^\pm},
\label{eq:ssw1} 
\end{equation} 
where $g^+ $ and $g^-$ are arbitrary functions of~$k_\parallel$, and where
\begin{equation}
n^+ + n^- = 6,
\label{eq:npnm} 
\end{equation} 
with $2<n^\pm < 4$.  Equations~(\ref{eq:ssw1}) and (\ref{eq:npnm})
match the results of weak turbulence theory for incompressible MHD
turbulence if one allows only for three-wave interactions among shear
Alfv\'en waves (Galtier \& Chandran 2006), or if one considers only
the limit that $k_\perp \gg k_\parallel$ (Galtier et al 2002).  If one
writes $n^\pm = 3 \pm \alpha$ with $|\alpha|<1$ and
sets $g^+(k_\parallel) = g^-(k_\parallel)$,
then equation~(\ref{eq:epsilon}) gives
\begin{equation}
\frac{\epsilon^+}{\epsilon^-} = \frac{2 + \alpha}{2-\alpha}.
\label{eq:fluxratio} 
\end{equation} 
In the limit~$\alpha \ll 1$, $\epsilon^+/\epsilon^- = 1+\alpha$,
in agreement with the weak-turbulence-theory result
for $k_\perp \gg |k_\parallel|$ (Lithwick \& Goldreich~2003),
as in the  weak-turbulence
advection-diffusion model of Lithwick \& Goldreich~(2003).

In steady state, $A^+(k_\perp,0)$ and $A^-(k_\perp,0)$ are forced to
be equal at the dissipation scale so that $\tau_k^+ = \tau_k^-$.  This
phenomenon of ``pinning'' was discovered by Grappin et al~(1983) for
strong MHD turbulence, and further elaborated upon by Lithwick \&
Goldreich (2003) for the case of weak turbulence.  The dominant
fluctuation type then has the steeper spectrum.  If
$\epsilon^+/\epsilon^-$ is fixed, then the ratio $w_{k_f}^+/w_{k_f}^-$
of the rms amplitudes of the two fluctuation types at the outer
scale~$k_f^{-1}$ increases as $k_d/k_f$ increases, where $k_d$ is the
dissipation wave number.  Alternatively, if $w_{k_f}^+/w_{k_f}^-$ is
fixed, then $\epsilon^+/\epsilon^-$ approaches unity as~$k_d/k_f
\rightarrow \infty$.

\begin{figure}[h]
\includegraphics[width=2.1in]{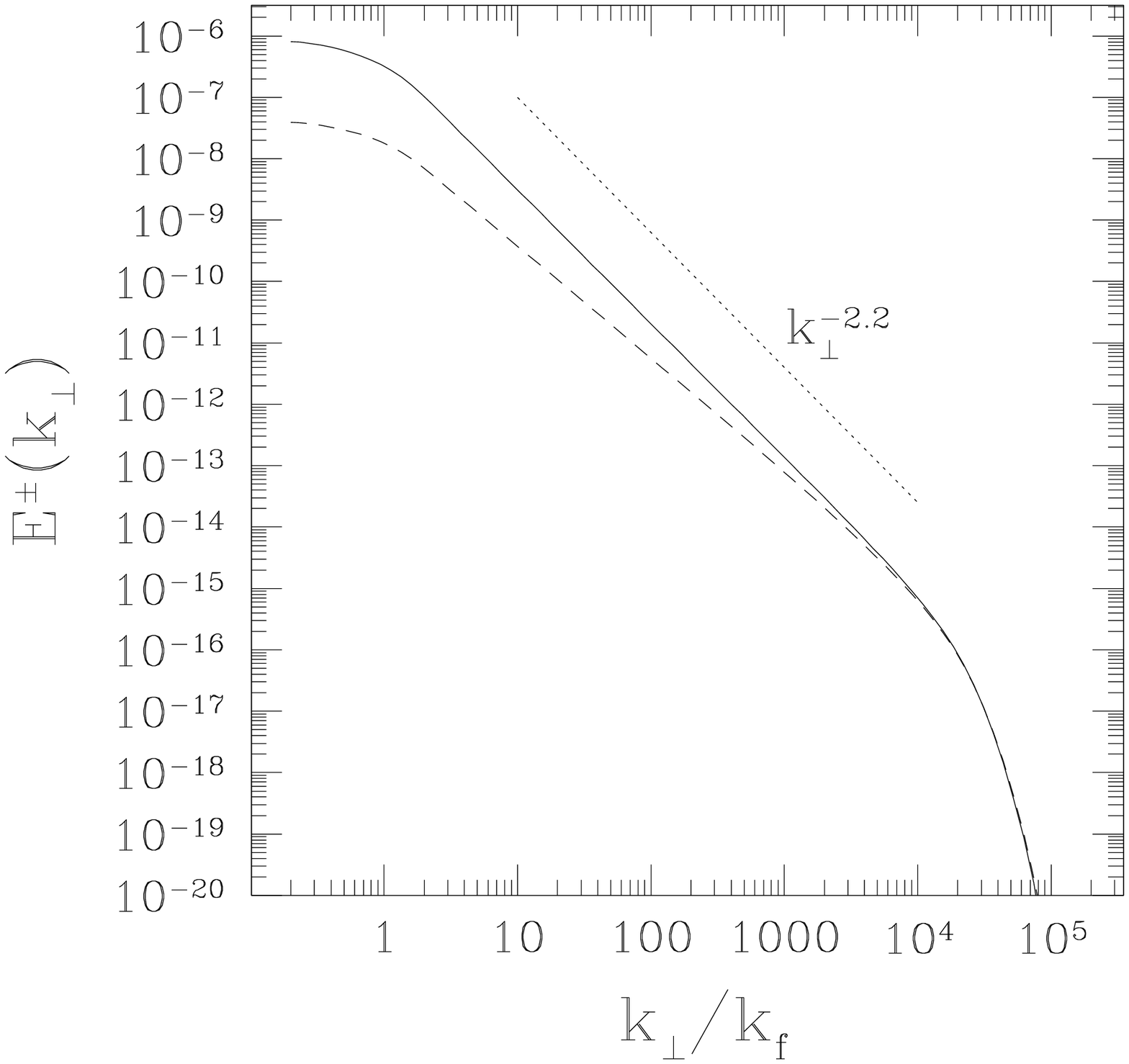}
\includegraphics[width=2.1in]{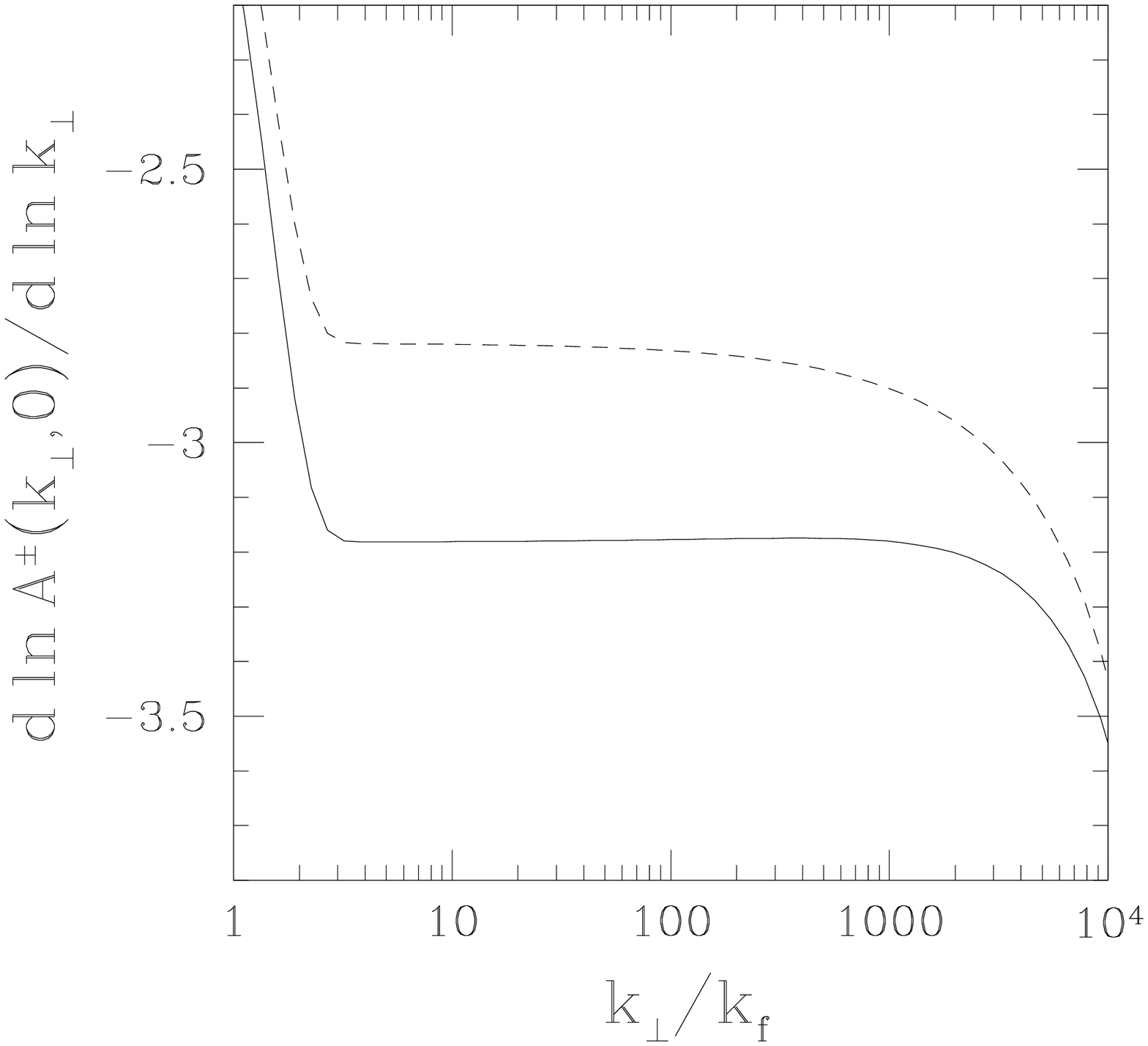}
\includegraphics[width=2.1in]{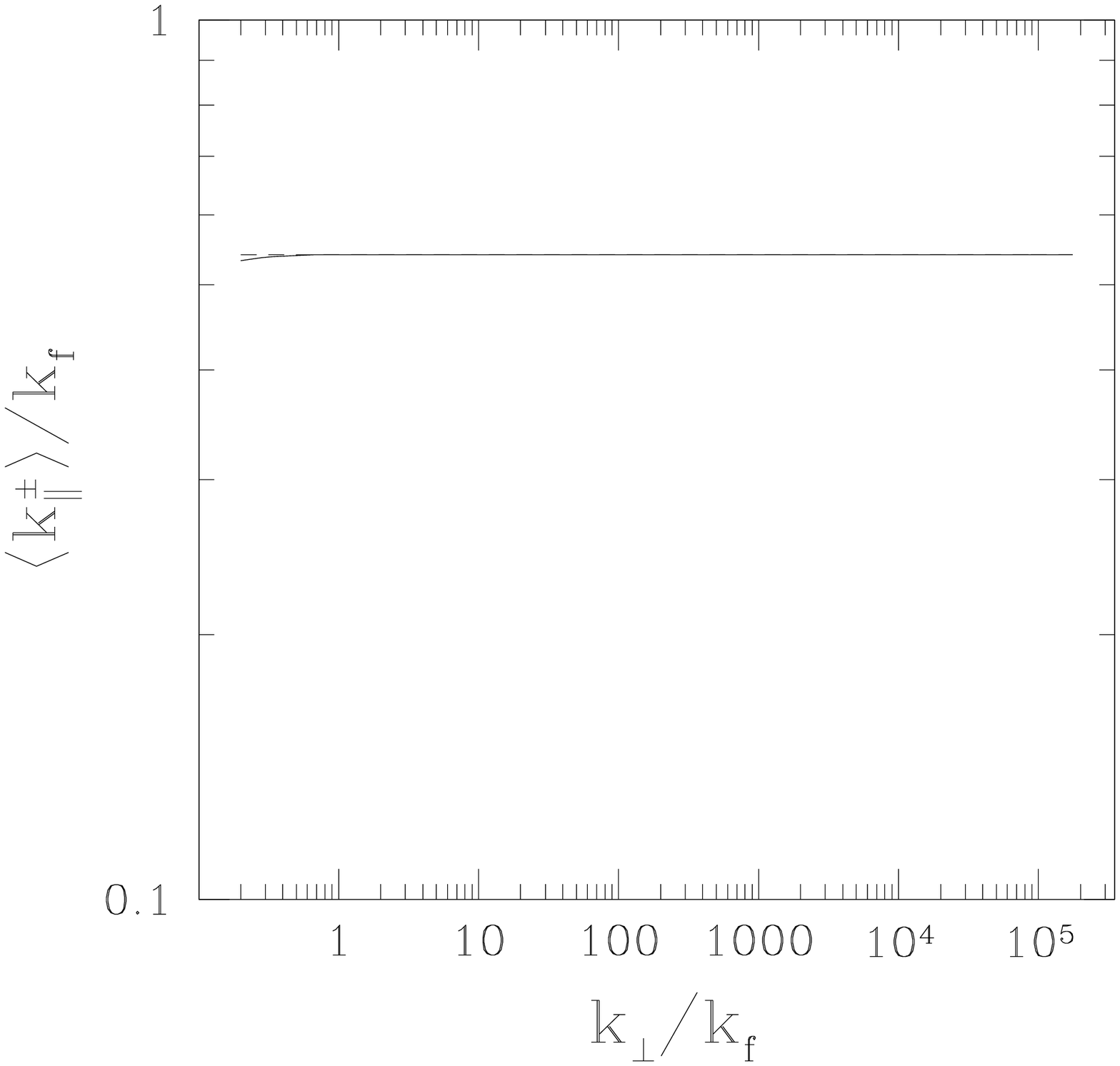}
\caption{\footnotesize 
Numerical solution of equation~(\ref{eq:FPpm})
in the weak-turbulence limit.
 {\em Left panel:} The dimensionless 1D
power spectrum defined in equation~(\ref{eq:defE}).
{\em Middle panel:} The spectral slopes at $k_\parallel=0$.
{\em Right panel:}
The weighted value of~$k_\parallel$ defined in equation~(\ref{eq:kparw}).
In all panels, the solid lines refer to~$w^+$ and the dashed
lines refer to~$w^-$. In the right-hand panel, the two lines
are almost on top of each other.
\label{fig:f2}}
\end{figure}

Several of these results are illustrated by the numerical solution to
equation~(\ref{eq:FPpm}) shown in Figure~\ref{fig:f2}. This 
solution is obtained using a logarithmic grid for $k_\perp$, with $k_{\perp,i} =
k_0 2^{i/n}$ for $0<i<N$.  Similarly, $k_{\parallel,j} = k_0 2^{j/n}$
for $1<j<M$, but $k_{\parallel,j} = 0$ for $j=0$.  $A^\pm_k$ is
advanced forward in time using a semi-implicit algorithm, in which the
terms $h_k^\pm$, $\tau_{{\rm eff},k_\perp}^\pm$, and $k_c^\pm$ on the
right-hand side of equation~(\ref{eq:FPpm}) are evaluated at the
beginning of the time step, and the $A^\pm$ terms on the right-hand
side of equation~(\ref{eq:FPpm}) are evaluated at the end of the time
step. The algorithm employs operator splitting, treating the $k_\perp$-advection,
forcing, and damping in one stage, and the $k_\parallel$-diffusion is
a second stage. In this approach, the matrix that has to be inverted
to execute each semi-implicit time step is tri-diagonal. An advantage
of this procedure over a fully explicit method is that the time step
is not limited by the $k_\parallel$-diffusion time at large~$k_\perp$
and small~$k_\parallel$. The discretized equations are written in
terms of the energy fluxes between neighboring cells, so that in the
absence of forcing and dissipation the algorithm conserves fluctuation
energy to machine accuracy. For the numerical solution plotted in
Figure~\ref{fig:f2}, $N=80$, $M=16$, $n=4$, $S^\pm = S_0^\pm k^2
\exp(- k^2/k_f^2)$, $S_0^+ = 1.2 S_0^-$, $k_f = 5 k_0$, and $\gamma^\pm_k
= 2k^2 \nu$, where $\nu$ is an effective viscosity. The
initial spectra are set equal to zero, and the equations are
integrated forward in time until a steady state is reached. In 
steady state, $\delta v^+ = 2.5\times 10^{-3} v_A$ and $\delta
v^- = 6.4 \times 10^{-4} v_A$.

The left-hand panel of Figure~\ref{fig:f2} is a plot of the dimensionless
one-dimensional power spectrum,
\begin{equation}
E^\pm(k_\perp) = \frac{k_0 k_\perp}{v_A^2} \int_{-\infty}^\infty dk_\parallel \, A^\pm(k_\perp,k_\parallel),
\label{eq:defE} 
\end{equation} 
which is proportional to the energy per unit~$k_\perp$ in $w^\pm$
fluctuations. The middle panel of Figure~\ref{fig:f2}  
shows that in the inertial range,
$d \ln A^+ (k_\perp,0)/d\ln k \simeq - 3.2$ and
$d \ln A^- (k_\perp,0)/d\ln k \simeq - 2.8$, as expected
for $S^+ /S^-  = 1.2$. The right-hand panel shows that the
weighted value of $k_\parallel$,
\begin{equation}
\langle k_\parallel^\pm \rangle = \frac{\displaystyle
  \int_{-\infty}^\infty dk_\parallel \, |k_\parallel|
  A^\pm(k_\perp,k_\parallel)}{\displaystyle \int_{-\infty}^\infty
  dk_\parallel \, A^\pm(k_\perp,k_\parallel)},
\label{eq:kparw} 
\end{equation} 
is roughly constant in the inertial range.

\section{Steady-State Strong Turbulence}
\label{sec:strong} 

This section addresses strong turbulence with $\chi_{k_f}^+ \sim 1$,
$\chi_{k_f}^- \la 1$, $w_{k_f}^+ \geq w_{k_f}^-$, and $\overline{
  k}_\parallel^{\,+} \sim \overline{ k}_\parallel^{\,-} $ at the outer
scale~$k_f^{-1}$.  The discussion allows for the possibility that
$\chi_{k_\perp}^- \ll 1$.  As fluctuation energy cascades to
larger~$k_\perp$, it diffuses to larger~$|k_\parallel|$, so that
$\overline{ k}_\parallel^{\,+}$ and $\overline{ k}_\parallel^{\,-}$
increase with increasing~$k_\perp$.  Moreover, for both $w^+$ and
$w^-$, the fluctuation energy diffuses over a $k_\parallel$-increment
of $\sim k_c^+$ during one cascade time.  For the steady-state
solutions of interest, $k_c^+$ is an increasing function of~$k_\perp$,
and thus at each $k_\perp$ we will have that $\overline{
  k}_\parallel^{\,+} \sim \overline{ k}_\parallel^{\,-} \sim
k_c^+$. One can thus define a single parallel-wavenumber, $\overline{
  k_\parallel}(k_\perp)$, to describe the spectra, with
\begin{equation}
 \overline{ k_\parallel} \sim
\overline{ k}_\parallel^{\,+} \sim \overline{ k}_\parallel^{\,-}
\sim k_c^+ 
\label{eq:defolkp} 
\end{equation} 
at each~$k_\perp$. Since $\overline{ k}_\parallel^{\,+}
\sim k_c^+$ at each scale, 
\begin{equation}
\chi_{k_\perp}^+ \sim 1
\label{eq:chistrong} 
\end{equation} 
throughout the inertial range. 
On the other hand, since $w_{k_\perp}^-$ can be much
less than~$w_{k_\perp}^+$, $\chi_{k_\perp}^-$ can be~$\ll 1$.

The cascade time for the $w^-$ fluctuations is given by the strong-turbulence
phenomenology of equation~(\ref{eq:tau2}), so that the energy
flux in $w^-$ fluctuations is
\begin{equation}
\epsilon^- \sim 
\frac{(w^-_{k_\perp})^2}{\tau_{k_\perp}^-}
\sim k_\perp w_{k_\perp}^+  (w_{k_\perp}^-)^2.
\label{eq:epsms} 
\end{equation} 
Allowing for the possibility that $\chi_{k_\perp}^- \ll 1$, the cascade
time of the $w^+$ fluctuations follows the weak-turbulence
phenomenology of equation~(\ref{eq:tau1}). This formula
is also accurate for $\chi_{k_\perp}^-$ as large as~1 (in which
case $w^-_{k_\perp} \sim w^+_{k_\perp}$). The energy
flux in $w^+$ fluctuations is then
\begin{equation}
\epsilon^+ \sim \frac{(w^+_{k_\perp})^2}{\tau_{k_\perp}^+}
\sim \frac{k_\perp^2 (w_{k_\perp}^-)^2 (w_{k_\perp}^+)^2 }{\overline{ k_\parallel} v_A}
\sim k_\perp w_{k_\perp}^+  (w_{k_\perp}^-)^2,
\label{eq:epsps} 
\end{equation} 
which is roughly the same as~$\epsilon^-$.  It is assumed that the
energy flux depends on the spectral slope as in weak turbulence, so
that the fluctuation type with the steeper spectrum has the larger
energy flux. If
\begin{equation}
w^\pm \propto k_\perp^{-a^\pm},
\label{eq:defapam} 
\end{equation} 
then equations~(\ref{eq:epsms}) and (\ref{eq:epsps})  imply that
when $\epsilon^+$ and $\epsilon^-$ are independent of~$k_\perp$,
\begin{equation}
a^+ + 2a^- = 1.
\label{eq:apam1} 
\end{equation} 
The condition that $\chi_{k_\perp}^+ \sim 1$ throughout the inertial range
then implies that
\begin{equation}
\overline{ k_\parallel} \propto k_\perp^{1- a^+}.
\label{eq:olkp1} 
\end{equation} 
As discussed in earlier studies (Grappin et~al~1983, Lithwick \&
Goldreich~2003), the spectra are pinned at the dissipation scale, so
that the dominant fluctuation type will have the steeper spectrum and
a somewhat larger energy flux.  For the zero-cross-helicity case,
equations~(\ref{eq:apam1}) and (\ref{eq:olkp1}) give $w_{k_\perp}^+ =
w_{k_\perp}^- \propto k_\perp^{-1/3}$ and $ \overline{ k}_\parallel
\propto k_\perp^{2/3}$, as in the work of Goldreich \& Sridhar (1995)
        [see also Higdon (1984)].

When $S_k^\pm = \gamma_k^\pm = 0$,
equation~(\ref{eq:FPpm})  possesses
an analytical solution that reproduces the above scalings.
This solution can be obtained by starting with the assumptions that
\begin{equation}
A^\pm (k_\perp,0) = c_3^\pm k_\perp^{-b^\pm},
\label{eq:ansatz} 
\end{equation} 
that the energy cascade is
dominated by local interactions, and that
$A^+(k_\perp,0) > A^-(k_\perp,0)$
for all~$k_\perp$. Equation~(\ref{eq:taueff}) then
becomes $(\tau_{{\rm eff},k_\perp}^\pm)^{-1} = k_\perp^4
A^\mp(k_\perp,0)/v_A$, and $k_{c,\rm max} = k_c^+$. Upon defining
\begin{equation}
f^\pm_k = k_\perp^{6 - b^\mp}A_{k}^\pm
\label{eq:deff} 
\end{equation} 
and
\begin{equation}
s = k_\perp^{8-2b^+},
\label{eq:defs} 
\end{equation} 
one can rewrite equation~(\ref{eq:FPpm}) as
\begin{equation}
\frac{\partial f^\pm_k }{\partial s} = D^\pm \frac{\partial ^2 f_k^\pm}{
\partial k_\parallel^2},
\label{eq:eqnf} 
\end{equation} 
with
\begin{equation}
D^\pm = \frac{c_2 (c_3^+)^2}{c_1 (8-2b^+)(b^\pm -1) v_A^4}.
\label{eq:defDpm} 
\end{equation} 
Equation~(\ref{eq:eqnf}) is solved by taking
\begin{equation}
f_k^\pm = \frac{c_4^\pm}{\sqrt{s}}\exp\left(- \frac{k_\parallel^2}{4D^\pm s}\right).
\label{eq:fsolve} 
\end{equation} 
Requiring that equation~(\ref{eq:ansatz}) 
be satisfied, one finds that $c_4^\pm = c_3^\pm$
and 
\begin{equation}
2b^+ + b^- = 10.
\label{eq:bpbm} 
\end{equation} 
The dominance of local interactions requires that $b^+ < 4$,
and thus $b^- > 2$. When forcing and dissipation are taken into account,
the exact solution becomes an approximate solution that is valid
only within the inertial range. In this case, $b^+ > b^-$ 
because the spectra are pinned at the dissipation
scale whereas $A^+(k_\perp,0)$ is larger
than~$A^-(k_\perp,0)$ within the inertial range.  
Equation~(\ref{eq:fsolve}) implies that
\begin{equation}
\overline{ k}_\parallel^{\,+} \simeq
\overline{ k}_\parallel^{\,-} \sim \sqrt{D^\pm s} \sim \frac{c_3^+ k_\perp^{4-b^+}}{v_A^2},
\label{eq:kpkm0} 
\end{equation} 
where  the dimensionless constants in the
expression for~$D^\pm$ have been dropped, but 
$c_3^+$, which has dimensions, has been kept.
Equations~(\ref{eq:defkc}), (\ref{eq:ansatz}), and~(\ref{eq:kpkm0}) 
show that $k_c^+ \sim \overline{ k}_\parallel^{\,+}$ for all $k_\perp$, so
that $\chi_{k_\perp}^+ \sim 1$ for all~$k_\perp$.
Equation~(\ref{eq:defE}) gives
\begin{equation}
E^\pm(k_\perp) \propto A^\pm(k_\perp,0) k_\perp \overline{ k}_\parallel^{\,+} \propto
k_\perp^{5-b^+ - b^\pm},
\label{eq:Eb} 
\end{equation} 
from which it follows that 
\begin{equation}
w_{k_\perp}^+ \propto k_\perp^{3-b^+},
\label{eq:wlps} 
\end{equation} 
and
\begin{equation}
w_{k_\perp}^- \propto k_\perp^{(6- b^+ - b^-)/2}.
\label{eq:wlms} 
\end{equation} 
This solution reduces to the critical-balance solution of Goldreich \&
Sridhar (1995) when $b^+ = b^- = 10/3$, in which case $\overline{
  k}_\parallel^{\, \pm} \propto k_\perp^{2/3}$, and $w_{k_\perp}^\pm \propto
k_\perp^{-1/3}$.  Comparing equations~(\ref{eq:wlps}) and
(\ref{eq:wlms}) with equation~(\ref{eq:defapam}), it can be seen that $b^+$
corresponds to $3+a^+$ and $b^-$ corresponds to $3 + 2a^- - a^+$.
Equation~(\ref{eq:kpkm0}) is thus equivalent to
equation~(\ref{eq:olkp1}), and equation~(\ref{eq:bpbm}) is equivalent
to equation~(\ref{eq:apam1}).

\begin{figure}[h]
\includegraphics[width=2.1in]{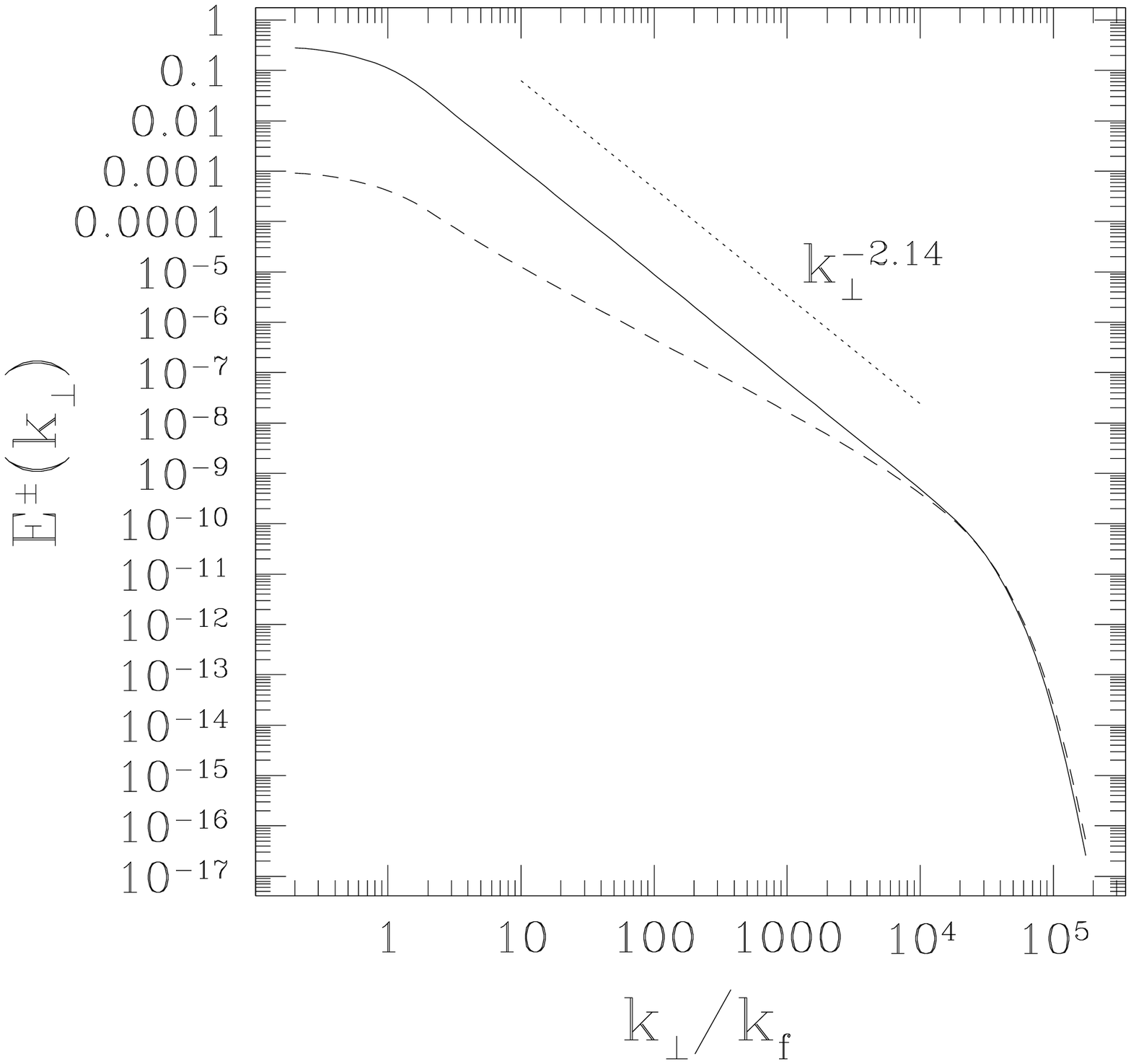}
\includegraphics[width=2.1in]{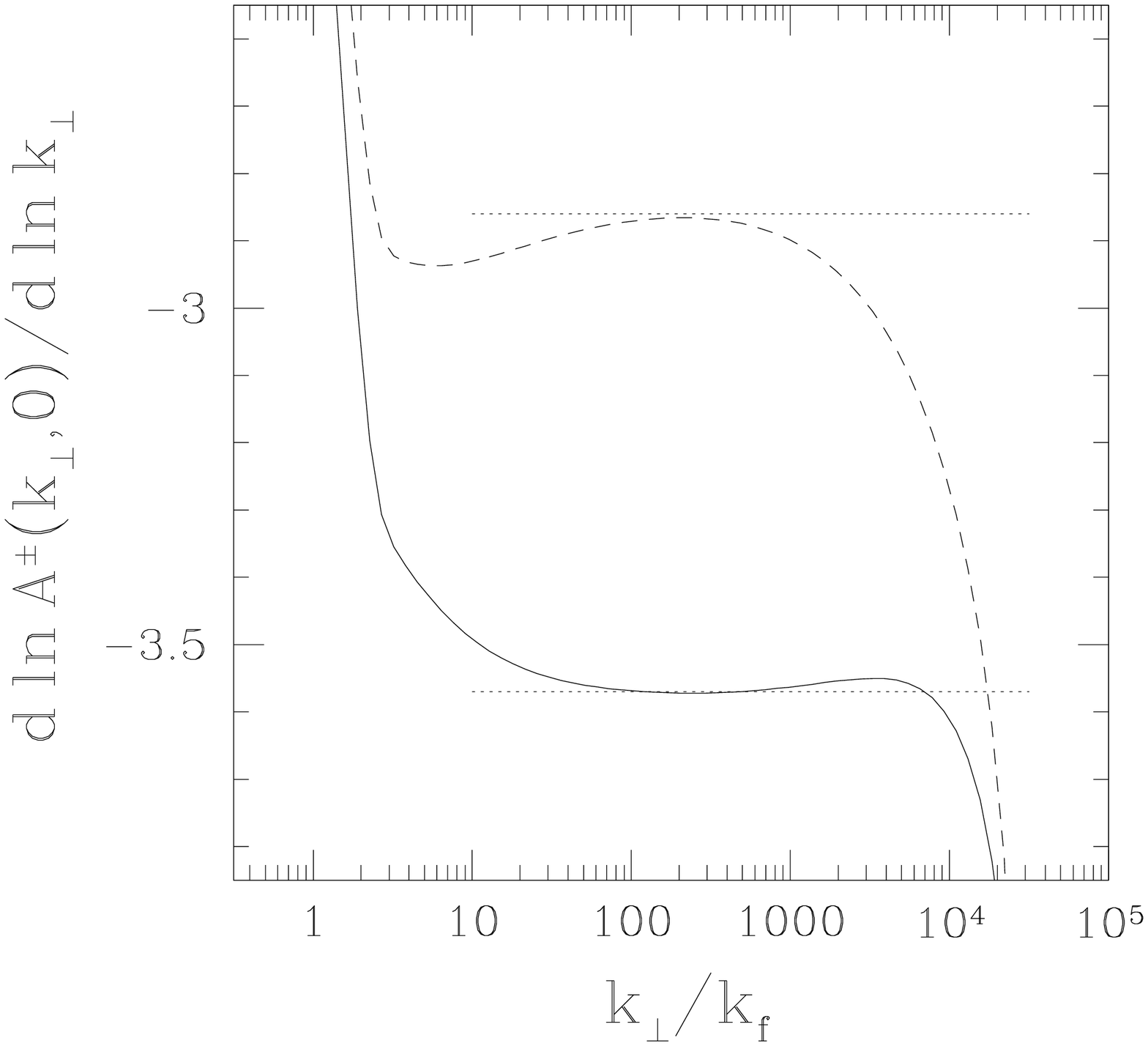}
\includegraphics[width=2.1in]{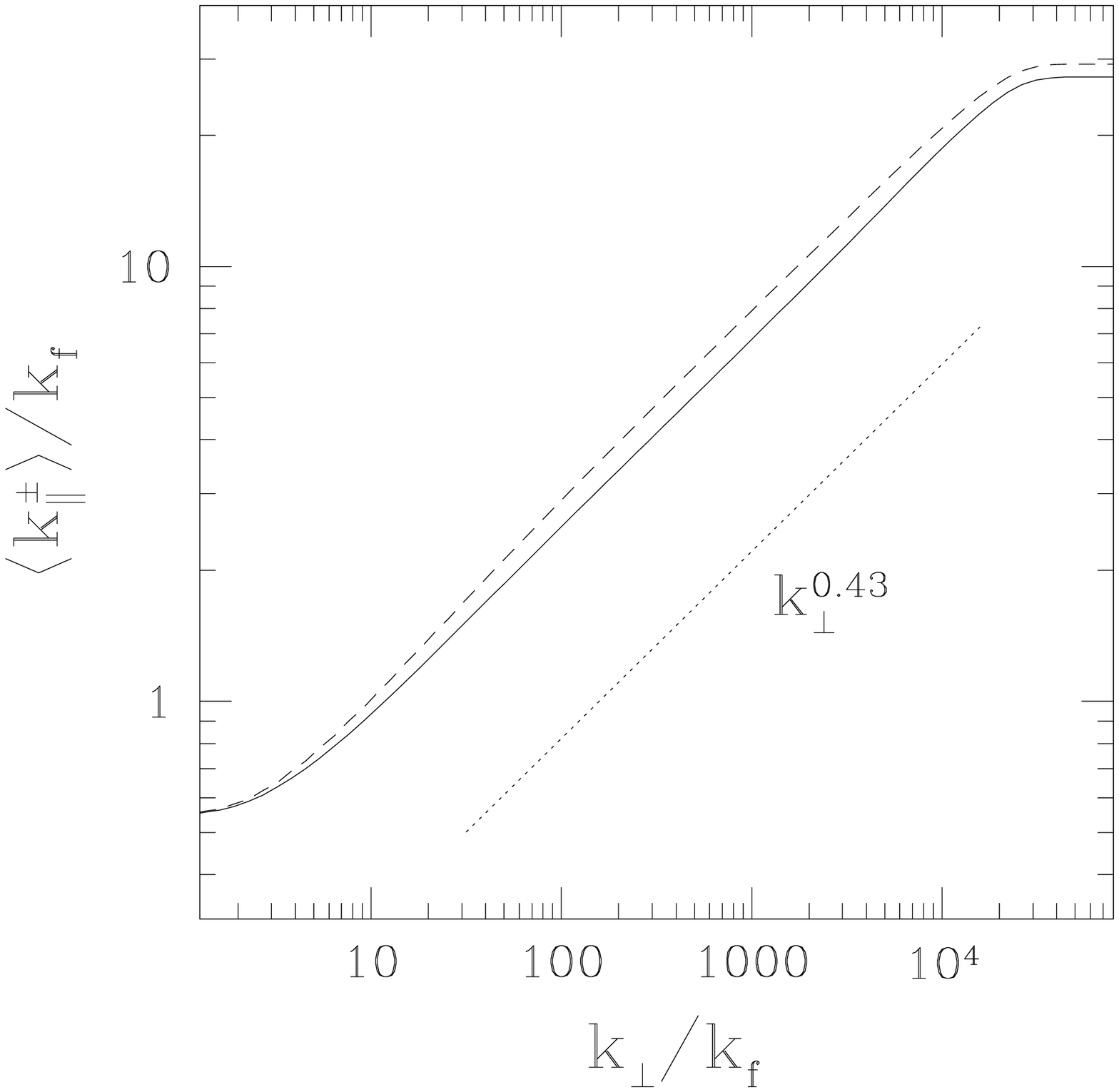}
\caption{\footnotesize   Numerical solution of equation~(\ref{eq:FPpm})
for strong turbulence with~$\chi_{k_\perp}^+\sim 1$.
{\em Left panel:} The dimensionless 1D
power spectrum defined in equation~(\ref{eq:defE}).
{\em Middle panel:} The spectral slopes at $k_\parallel=0$.
{\em Right panel:}
The weighted value of~$k_\parallel$ defined in equation~(\ref{eq:kparw}).
In all panels, the solid lines refer to~$w^+$ and the dashed
lines refer to~$w^-$.
\label{fig:f3}}
\end{figure}

Figure~\ref{fig:f3} shows the results from a numerical solution of
equation~(\ref{eq:FPpm}), obtained by integrating forward in time as
described in section~\ref{sec:weak} with the spectra initially equal
to zero. The numerical solution was obtained by setting $S^\pm = S_0^\pm
k^2 \exp(- k^2/k_f^2)$ with $S_0^+ =
1.2 S_0^-$ and using the parameters
(defined in section~\ref{sec:weak}) $N=80$, $M=60$,
$n=4$, $k_f= 5 k_0$,   and $\gamma_k^\pm = 2k^2 \nu$, where the constant $\nu$ is
an effective viscosity. The rms velocities at steady
state are $\delta v^+ = 1.5 v_A$ and $\delta v^- = 0.10 v_A$.  The
left-hand panel shows that the one-dimensional energy
spectrum~$E^+_{k_\perp}$ is $\propto k_\perp^{-2.14}$ in the inertial
range, which corresponds to~$b^+ = 3.57$ in equation~(\ref{eq:Eb}).
Equation~(\ref{eq:bpbm}) then gives $b^- = 2.86$. The dotted lines in
the middle panel of Figure~\ref{fig:f3} correspond to the values of
$b^+ = 3.57$ and $b^- = 2.86$, which are reasonably close to the
values of $- d \ln A^\pm(k_\perp,0)/d\ln k_\perp$ in the numerical
solution, although these latter values vary throughout the inertial
range in the numerical solution.  For $b^+ = 3.57$,
equation~(\ref{eq:kpkm0}) gives $\overline{ k}_\parallel^{\,+}\propto
k_\perp^{0.43}$, which is a close match to the numerical solution, as
shown in the right-hand panel of Figure~\ref{fig:f3}.  The left-hand
panel of Figure~\ref{fig:f3} shows that the steady-state solutions for
$A^+$ and $A^-$ are ``pinned'' at the dissipation scale, as expected.

It should be noted that when $\chi_{k_\perp}^+ \sim 1$ and
$\chi_{k_\perp}^- \ll 1$, the dominant~$w^+$ fluctuations are only
weakly damped by nonlinear interactions with $w^-$ waves, in the sense
that $\tau_{k_\perp}^+$ is much larger than the linear wave period.
On the other hand, for the smaller-amplitude $w^-$
fluctuations, the linear wave period and cascade time are comparable.
Thus, paradoxically, the larger-amplitude~$w^+$ fluctuations
can be described as waves, or, more precisely, a non-sinusoidal
wave train, whereas the smaller-amplitude $w^-$ fluctuations
can not be accurately described as waves.

\section{Transition Between Weak Turbulence and Strong Turbulence}
\label{sec:trans} 

\begin{figure}[h]
\includegraphics[width=2.1in]{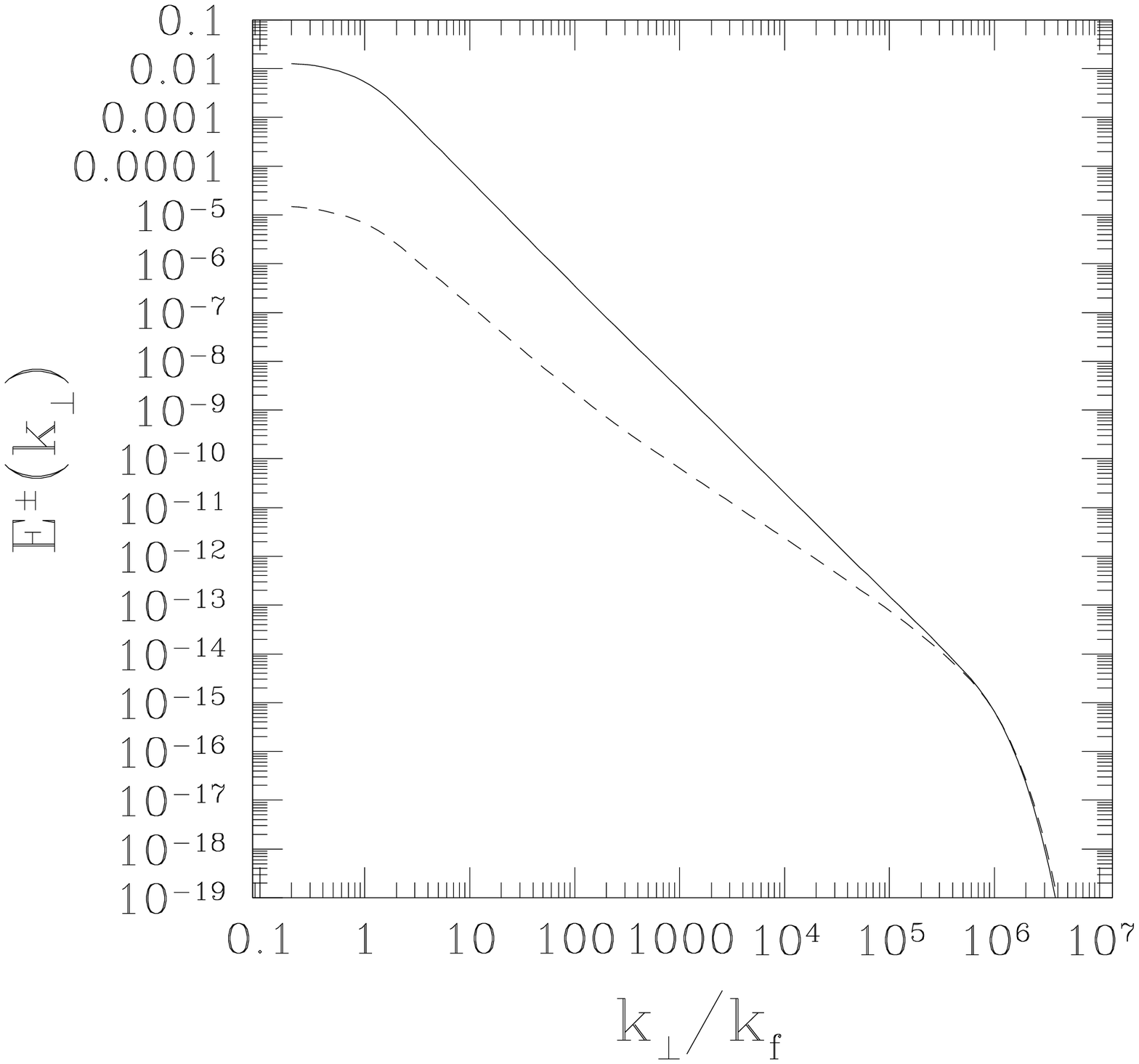}
\includegraphics[width=2.1in]{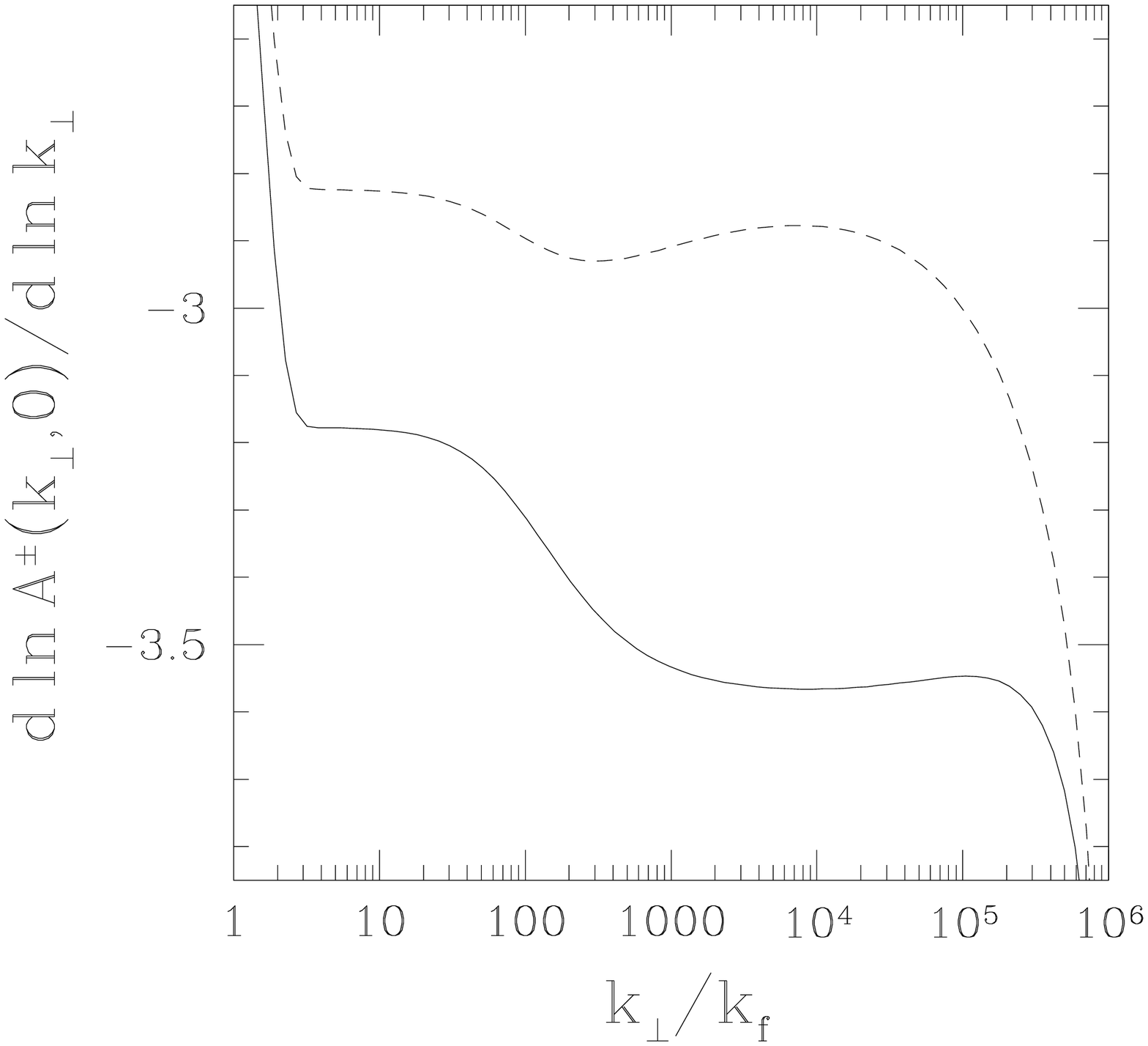}
\includegraphics[width=2.1in]{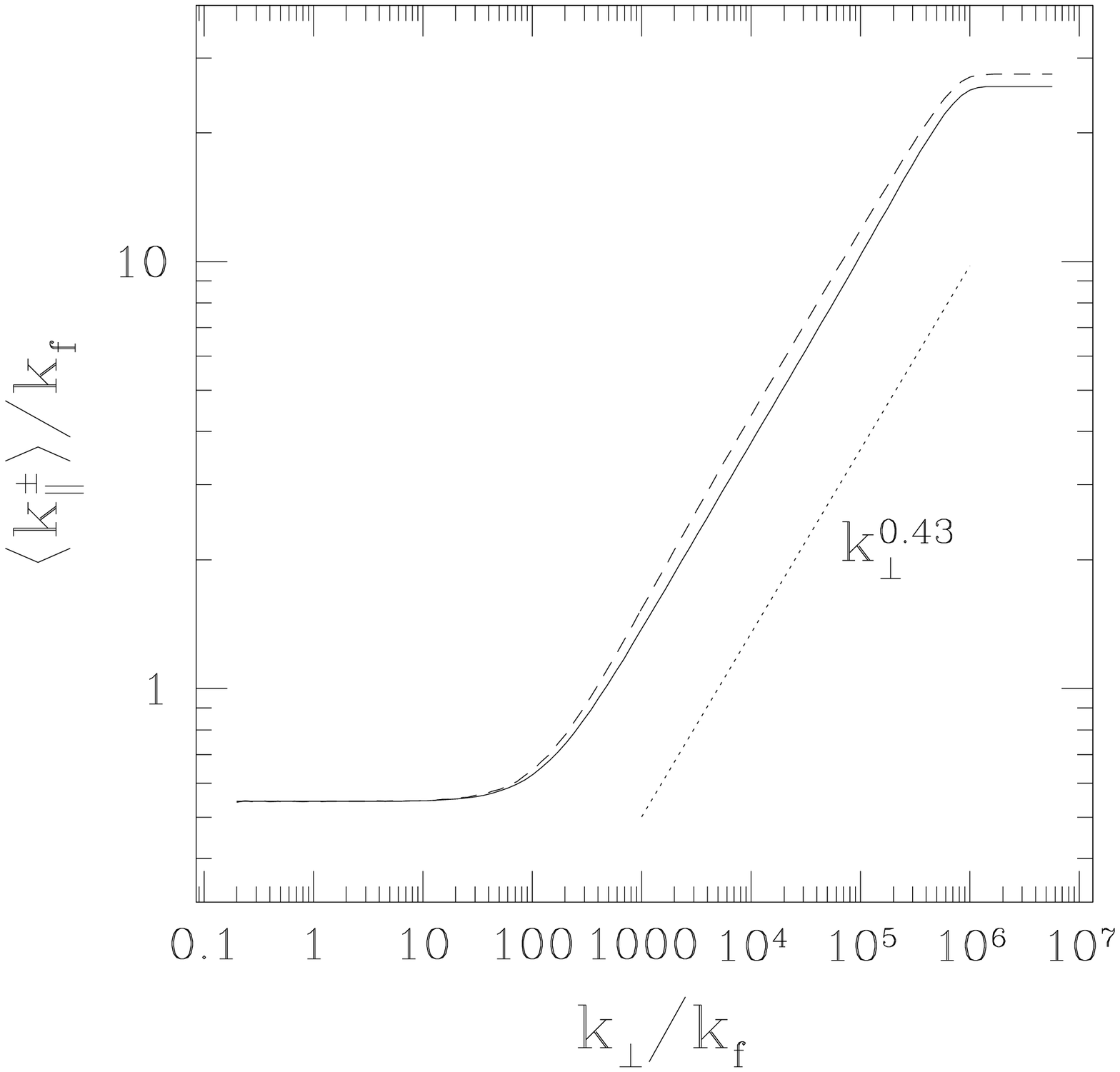}
\caption{\footnotesize   Numerical solution of equation~(\ref{eq:FPpm})
showing a smooth transition from the
weak-turbulence limit at small~$k_\perp$ to the strong-turbulence
limit at large~$k_\perp$. 
{\em Left panel:} The dimensionless 1D
power spectrum defined in equation~(\ref{eq:defE}).
{\em Middle panel:} The spectral slopes at $k_\parallel=0$.
{\em Right panel:}
The weighted value of~$k_\parallel$ defined in equation~(\ref{eq:kparw}).
In all panels, the solid lines refer to~$w^+$ and the dashed
lines refer to~$w^-$. 
\label{fig:f4}}
\end{figure}

This section again addresses turbulence in which
$\overline{k}_\parallel^{\,+} \sim\overline{k}_\parallel^{\,-} $ at
the outer-scale wavenumber,~$k_f$.  In the weak-turbulence limit,
$\chi_{k_\perp}^+$ and~$\chi_{k_\perp}^-$ increase with
increasing~$k_\perp$.  If the dissipation wavenumber~$k_d$ is
sufficiently large, then $\chi_{k_\perp}^+$ and/or $\chi_{k_\perp}^-$
will increase to a value of order unity at some $k_\perp$ within the
inertial range. This perpendicular wavenumber is denoted~$k_{\rm
  trans}$. The turbulence will then be described by the
weak-turbulence scalings of section~\ref{sec:weak} for~$k_f\ll k_\perp
\ll k_{\rm trans}$, and by the strong-turbulence scalings of
section~\ref{sec:strong} for $k_{\rm trans} \ll k_\perp \ll k_d$.
Figure~\ref{fig:f4} shows a numerical solution of
equation~(\ref{eq:FPpm}) that illustrates how the turbulence makes
this transition in a smooth manner. At small wavenumbers, this
solution is similar to the weak-turbulence solution plotted in
Figure~\ref{fig:f2}, and at large wavenumbers it is similar to the
strong-turbulence solution plotted in Figure~\ref{fig:f3}.  The
solution shown in Figure~\ref{fig:f4} was obtained by integrating
equation~(\ref{eq:FPpm}) forward in time to steady state using the
numerical method described in section~\ref{sec:weak}. The spectra were
initially set equal to zero. The numerical solution was obtained by
setting $S^\pm = S_0^\pm k^2 \exp(- k^2/k_f^2)$ with $S_0^+ = 1.2
S_0^-$ and using the parameters $N=100$, $M=56$, $n=4$, $k_f= 5 k_0$,
and $\gamma_k^\pm = 2k^2 \nu$, where the constant $\nu$ is an
effective viscosity.  The rms velocities at steady state are $\delta
v^+ = 0.32 v_A$ and $\delta v^- = 0.012 v_A$.

\section{Unequal Parallel Correlation Lengths at the Outer Scale}
\label{sec:unequal} 

In sections~\ref{sec:weak} through \ref{sec:trans}, it was assumed
that $\overline{ k}_\parallel^{\, + } \sim \overline{ k}_\parallel^{\,
  - }$ at the outer scale. This assumption is applicable to many
settings. For example, in a plasma of dimension~$L$ that is stirred by
a force that has a correlation length~$l \ll L$, the velocity
fluctuations that are excited have a correlation length~$l$, and this
correlation length is imprinted on both the~$w^+$ and $w^-$
fluctuations. On the other hand, if waves are launched along the
magnetic field into a bounded plasma from opposite sides of the
plasma, and the waves from one side have a much larger parallel
correlation length than the waves from the other side, it is
possible to set up turbulence in which the two wave types have very
different parallel correlation lengths at the outer scale.  This
situation is discussed briefly in this section.

For strong turbulence, if both $\chi_{k_\perp}^+$ and
$\chi_{k_\perp}^-$ are $\sim 1$ at some perpendicular
scale~$k_\perp^{-1}$, but one fluctuation type, say $w^+$, has a much
smaller parallel correlation length than the other (and thus a much
larger amplitude), then during a time~$\tau_{k_\perp}^-$ the
``propagation with distortion'' mechanism discussed in
section~\ref{sec:parcasc} will increase $\overline{
  k}_\parallel^{\,-}$ until it equals $\overline{ k}_\parallel^{\,+}$,
which will cause $\chi_{k_\perp}^-$ to become~$\ll 1$ at
scale~$k_\perp^{-1}$.  At smaller scales, the solution can be
described by the scalings presented in section~\ref{sec:strong}, in
which $\overline{ k}_\parallel^{\,+}(k_\perp) \sim \overline{
  k}_\parallel^{\,-}(k_\perp)$. Similarly, if $\chi_{k_\perp}^+ \sim
1$ but $\chi_{k_\perp}^- \ll 1$ at some scale $k_\perp^{-1}$ and if
$\overline{ k}_\parallel^{\,+} \gg \overline{ k}_\parallel^{\,-}$ at
that scale, then during a time~$\tau_{k_\perp}^-$ the ``propagation
with distortion'' mechanism discussed in section~\ref{sec:parcasc}
will again increase $\overline{ k}_\parallel^{\,-}$ until it equals
$\overline{ k}_\parallel^{\,+}$, the parallel scales will remain
comparable at smaller perpendicular scales, and the solution can be
described by the scalings in section~\ref{sec:strong}.  The case in
which $ \chi_{k_\perp}^+ \sim 1$, $\chi_{k_\perp}^- \ll 1$, and
$\overline{ k}_\parallel^{\,+} \ll \overline{ k} _\parallel^{\,-}$ is
not addressed in this paper.

\section{Implications for Turbulence in the Solar Corona and Solar Wind}
\label{sec:solarwind} 

In this section, the preceding analysis of incompressible MHD
turbulence is applied to the solar wind and solar corona. It should be
noted at the outset, however, that the solar wind and solar corona
(beyond roughly~$r=1.5R_{\sun}$, where $r$ is distance from the Sun's
center) are in the collisionless regime, and the pressure tensor is
not isotropic as assumed in ideal MHD. Moreover, the value of $\beta =
8\pi p/B^2$ is $\ll 1$ in the corona and typically $\sim 1$ in the
solar wind at 1~AU, whereas incompressible MHD corresponds to the
limit~$\beta \rightarrow \infty$. A preliminary question that needs to
be addressed is thus the extent to which incompressible MHD is an
accurate model for these plasmas.

Schekochihin et~al~(2007) have recently carried out extensive
calculations based on kinetic theory that provide a detailed answer to
this question. These authors examined anisotropic turbulence in weakly
collisional magnetized plasmas using gyrokinetics, a low-frequency
expansion of the Vlasov equation that averages over the gyromotion of
the particles.  By applying the form of the gyrokinetic expansion
derived by Howes et~al~(2006), Schekochihin et~al~(2007) showed
analytically that non-compressive Alfv\'enic turbulence in the
quasi-2D regime (i.e., $k_\perp \gg k_\parallel$) can be accurately
described using reduced MHD in both the collisional and collisionless
limits, regardless of~$\beta$, provided that the length scales of the
fluctuations are much larger than the proton gyroradius and the
frequencies are much less than the proton cyclotron frequency.  Since
non-compressive quasi-2D fluctuations are thought to be the dominant
component of the turbulence in the solar wind (see, e.g., Bieber
et~al~1994) and the solar corona (Dmitruk \& Matthaeus~2003, Cranmer
\& van Ballegooijen~2005), incompressible MHD is a useful
approximation for modeling turbulence in these settings.

\subsection{Cross helicity in the solar wind and solar corona}
\label{sec:CH} 

Cross helicity in the solar wind has been measured {\em in situ} by
several different spacecraft.  In terms of the Elsasser variables
$w^\pm$, there is a substantial excess of outward propagating
fluctuations (taken to be $w^+$ throughout this section) over inward
propagating fluctuations (taken to be $w^-$) in the inner heliosphere,
although this imbalance decreases with increasing~$r$, as seen, for
example, in Voyager data for low heliographic latitude (Matthaeus \&
Goldstein 1982, Roberts et~al~1987) and Ulysses data at high latitude
(Goldstein et~al~1995).  In a study of Ulysses and Helios data,
Bavassano et~al~(2000) found that $e^+/e^- \propto r^{-1.02}$ for
$r<2.6$~AU and $e^+/e^- \sim 2$ for $3\mbox{ AU}\la r \la 5 \mbox{
  AU}$, where $e^\pm$ is the energy per unit mass associated with
$w^\pm$ fluctuations. These numbers are intended as illustrative
average values, as individual measurements of $e^+/e^-$ in the solar
wind vary significantly.

Although it has not been directly measured, the ratio $e^+/e^-$ is
likely very large in open-field-line regions of the solar corona.
This can be seen from the work of Cranmer \& van Ballegooijen (2005),
who modeled the generation of Alfv\'en waves by the observed motions
of field-line footpoints in the photosphere, and the propagation and
reflection of these waves as they travel along open field lines from
the photosphere out into the interplanetary medium. They found that
the ratio of the frequency-integrated rms Elsasser variables ($w^+$
and $w^-$) is $\sim 30$ at $r= 2R_{\sun}$ (i.e., $e^+/e^- \sim 900$).
Verdini \& Velli~(2007) developed a different model for the
propagation, reflection, and turbulent dissipation of Alfv\'en waves
in the solar atmosphere and solar wind and found that $e^+/e^-\simeq
80$ at $r=2R_{\sun}$. Based on these results, one can make the rough
estimate that
\begin{equation}
\frac{w_{k_f}^+}{w_{k_f}^-} \sim 10 \hspace{1cm} \mbox{ (at $r=2R_{\sun}$),}
\label{eq:wpwmra} 
\end{equation} 
where $k_f$ is the perpendicular wavenumber at the outer scale.

\subsection{Is quasi-2D turbulence in the corona and
solar wind weak or strong?}
\label{sec:q2dws} 

In much of the solar wind, $\delta B$ is comparable to $B_0$, and the
turbulence is in the strong-turbulence regime with~$\chi_{k_f}^+ \sim
1$.  For the corona, Cranmer \& van Ballegooijen (2005) found that the
outer-scale fluctuations in open-field-line regions have periods of
$T= 1-5$ minutes, $\delta v \sim 100$~km/s, and perpendicular
correlation lengths of~$L_\perp \sim k_f^{-1} \sim 10^4$~km.  The
Alfv\'en speed in their model corona is between 2000 and 3000~km/s at
$r=2R_{\sun}$, and thus $\delta v \ll v_A$.  However, the parallel
correlation length $L_\parallel$ of the outer-scale fluctuations is
$\sim v_A T = 1.8 -9 \times 10^5$~km, which is $\gg L_\perp$. Because
$L_\parallel/L_\perp \sim v_A/\delta v$,
\begin{equation}
\chi_{k_f}^+ \sim 1
\label{eq:chicorona} 
\end{equation} 
and the low-frequency fluctuations launched into
the corona by footpoint motions are in the strong-turbulence regime.
There may be an additional population of higher-frequency waves in the
weak-turbulence regime, but these are not discussed here.

\subsection{Parallel correlation lengths of inward and outward waves}
\label{sec:pcl} 

In open-field-line regions of the corona, when $w^+$ waves are
reflected, the resulting $w^-$ waves have the same frequencies as the
$w^+$ waves. On the other hand, wave-reflection is more efficient at
lower frequencies (Velli 1993), so if there is a range of wave
frequencies at each~$k_\perp$, the energy-weighted average frequency
of inward waves would tend to be somewhat lower than that of the
outward waves. This suggests that at the outer scale the parallel
correlation length~$L_\parallel$ of the $w^-$ fluctuations is somewhat
larger than the value of~$L_\parallel$ for the $w^+$ fluctuations.
However, given equation~(\ref{eq:chicorona}), the $w^+$ fluctuations
imprint their parallel correlation length on the $w^-$ fluctuations
during a single turnover time $\tau_{k_\perp}^-$, as argued in
section~\ref{sec:unequal}. The parallel correlation lengths of the
$w^+$ and $w^-$ fluctuations in the corona can thus be taken to be
approximately equal at the outer scale, and hence also at smaller
scales. The same approximation is reasonable for turbulence in the
solar wind.

\subsection{Energy Dissipation Rate}
\label{sec:epsilon} 

If we take $k_f$ to be the perpendicular wave number at the outer
scale, $w_{k_f}^+$ to be the rms amplitude of the outward-propagating
fluctuations at the outer scale, and $w_{k_f}^-$ to be the rms
amplitude of the Sunward-propagating fluctuations at the outer scale,
then equations~(\ref{eq:epsms}), (\ref{eq:epsps}),
and~(\ref{eq:chicorona}) imply that
\begin{equation}
\epsilon^+ \sim \epsilon^- \sim  k_f w^+_{k_f} (w^-_{k_f})^2.
\label{eq:epsSW} 
\end{equation} 
This estimate is a factor of $\sim w_{k_f}^-/w_{k_f}^+$ smaller than
the standard strong-turbulence estimate of $\epsilon^+ \sim k_f
(w_{k_f}^+)^2 w_{k_f}^-$ that appears in many studies (e.g., Zhou \&
Matthaeus 1990, Cranmer \& Van Ballegooijen~2005, Lithwick, Goldreich,
\& Sridhar~2007, Verdini \& Velli~2007). This difference has important
implications for turbulent heating of the solar corona and solar wind.

\subsection{Cascade Time}
\label{sec:cascadetime}

For the energetically dominant $w^+$ fluctuations, the cascade
time~$\tau_{k_\perp}^+$ is much longer than the linear wave period, at
least at scales much larger than the dissipation scale.  This result
is important for determining the conditions under which turbulence can
be a viable mechanism to explain the heating of the solar
corona. Observations taken with the {\em Ultraviolet Coronagraph
  Spectrometer} (UVCS) indicate that there is strong heating of
coronal plasma at $r\lesssim 2 R_{\sun}$ (Kohl et al 1998, Antonucci
et al 2000). An appealing model to explain this heating is that
low-frequency Alfv\'en waves are launched by turbulent motions of
field-line footpoints in the photosphere, that some of these waves are
reflected, and that interactions between oppositely directed Alfv\'en
wave packets in the corona causes the wave energy to cascade to small
scales and dissipate (Matthaeus et~al~1999, 2002; Dmitruk et~al~2001,
2002; Cranmer \& van Ballegooijen 2005, 2007; Verdini \& Velli 2007).
In one version of this model, the waves that cross the transition
region and enter the corona have not yet undergone a turbulent
cascade, and their energy is concentrated at the fairly long periods
($>1$~minute) characteristic of the observed footpoint motions
that are believed to make the dominant contribution to the outward
directed wave flux.  In order for this scenario to explain the UVCS
measurements, there needs to be time for the outer-scale waves to
cascade within the corona before they travel beyond~$r\simeq 2
R_{\sun}$.  If, as above, we take $k_f$ to be the value of~$k_\perp$
at the outer scale, $w_{k_f}^+$ to be the rms amplitude of the outward
waves at the outer scale, and $L_\parallel$ to be the parallel
correlation length of the fluctuations at the outer scale, then
equation~(\ref{eq:tau1}) can be used to express the cascade time for
the outward waves at the outer scale as
\begin{equation}
\tau^+_{k_f} \sim \frac{L_\parallel}{v_A} \left(\frac{w_{k_f}^+}{w_{k_f}^-}\right)^2
(\chi_{k_f}^+)^{-2},
\label{eq:taucorona} 
\end{equation} 
where $\chi_{k_f}^+ \sim k_f w_{k_f}^+ L_\parallel/v_A$.  Thus, for
waves with a period $L_\parallel/v_A \sim 1$~minute,
equations~(\ref{eq:wpwmra}), (\ref{eq:chicorona}),
and~(\ref{eq:taucorona}) give $\tau^+_{k_f} \sim 100$~minutes. On the
other hand, the Alfv\'en speed in a coronal hole at~$r< 2R_{\sun}$
is~$\sim 2000-3000$~km/s (Cranmer \& van Ballegooijen 2005), and the
time for an Alfv\'en wave to travel from the coronal base out to
$r=2R_{\sun}$ is 4-6 minutes.  There is thus not enough time for the
energy of waves with periods $> 1$~minute to cascade and dissipate
within a few solar radii of the Sun.

Dmitruk \& Matthaeus~(2003) and Verdini \& Velli (2007) avoid this
difficulty by postulating that a broad frequency spectrum of waves is
launched upwards from the photosphere. Another possible way around
this difficulty is the development of a broad frequency spectrum of
fluctuations from wave-packet collisions in the chromosphere, in which
the energies of inward and outward propagating waves are
comparable due to strong wave reflection at the transition region.

\subsection{Spectral Index}
\label{sec:slope} 

Much of the discussion of the inertial-range power spectrum of
solar-wind turbulence has focused on the question of whether the
spectral index is closer to the Kolmogorov (1941) value of~$-5/3$ or
the Iroshnikov-Kraichnan value of~$-3/2$ (Iroshnikov~1963,
Kraichnan~1965).  A value of $-5/3$ is supported by a number of
theoretical studies (e.g., Montgomery \& Turner 1981, Higdon 1984,
Goldreich \& Sridhar 1995) and numerical simulations (Cho \& Vishniac
2000, M\"uller \& Biskamp 2000, Cho et~al~ 2002, Haugen et~al~2004). A
value value of~$-3/2$ is supported by a second group of theoretical
studies (Boldyrev 2005, 2006; Mason et~al~2006; see also Beresnyak \&
Lazarian 2006) and numerical simulations (Maron \& Goldreich 2001,
M\"uller et~al~2003, M\"uller \& Grappin 2005, Mininni \&
Pouquet~2007).  It should be noted that all of the above-mentioned
studies address MHD turbulence with negligible cross helicity.

Spacecraft measurements yield frequency spectra for the magnetic field
and velocity fluctuations, where the frequency $f$ is approximately
$k_r U/2\pi$, where $k_r$ is the radial component of the wave-vector
and $U$ is the solar-wind speed [Taylor's (1938) ``frozen-in flow
  hypothesis'']. Below a spectral-break frequency~$f_b$, the spectra
are typically fairly flat, being approximately proportional
to~$f^{-1}$ (Matthaeus \& Goldstein 1986). At $f>f_b$, the spectra
steepen. The time scale corresponding to the spectral break,
$f_b^{-1}$, increases with increasing~$r$.  For example, Bruno \&
Carbone (2005) found that $f_b^{-1}$ was $0.06$~hours at 0.3~AU,
0.16~hours at 0.7~AU, and 0.4~hours at 0.9~AU in a sample of Helios~2
data. In two other studies based on data from several spacecraft,
Matthaeus \& Goldstein (1986) found that $f_b^{-1}\sim 3.5$~hours at
1~AU, while Klein et~al~(1992) found $f_b^{-1} \sim 12$~hours at
4~AU. The inertial range roughly corresponds to frequencies in the
interval $f_b < f < f_d$, where~$f_d$ is the frequency corresponding
to the dissipation scale.  At 1~AU, $f_d \sim 0.3 \mbox{ s}^{-1}$
(Smith et~al~2006).  A large number of inertial-range spectral indices
have been reported in the literature.  For example, Matthaeus \&
Goldstein (1982) found a spectral index of $-1.73 \pm 0.08$ for the
magnetic field in Voyager data at $r=1$~AU, and a spectral index of
$-1.69 \pm 0.08$ for the total energy.  Goldstein et~al~(1995, Fig. 1)
found that the spectral index for the $w^+$ fluctuations was slightly
steeper than $-5/3$ in Ulysses data at 2~AU and 4~AU. Their results
also suggest a shallower $w^-$ spectrum, consistent with the idea that
the spectra are pinned at the dissipation wavenumber~$k_d$.  In a
study of Helios~2 magnetic-field data, Bruno \& Carbone (2005, Figure
23) found spectral indices of $-1.72$ at 0.3~AU, $-1.67$ at 0.7~AU,
and $-1.70$ at 0.9~AU.  Using data from the WIND spacecraft at 1~AU,
Podesta et~al~(2007) found a total-energy spectral index of $-1.63$,
with the velocity spectrum flatter than the magnetic spectrum.  Marsch
\& Tu (1996) found a spectral index of $-1.65\pm 0.01$ for the
magnetic field at 1~AU in Helios~2 data.  Horbury \& Balogh (1995)
found a spectral index close to~$-5/3$ for the magnetic field in
Ulysses data at 2.5~AU.  Using magnetic-field data from the ACE
spacecraft at 1~AU, Smith et~al~(2006) found a spectral index of
$-1.63\pm 0.14$ in open-field-line regions and $-1.56\pm 0.16$ in
magnetic clouds.  Smith (2003) found spectral indices between $-1.7$
and $-1.8$ in a study of Ulysses magnetic-field data covering a range
of heliographic latitudes and radii.

Overall, the spectra are more consistent with a Kolmogorov scaling
than an Iroshnikov-Kraichnan scaling. It should be emphasized,
however, that the observations in several cases are consistent with
inertial range spectra that are steeper than a Kolmogorov spectrum.
Spectral indices~$> 5/3$ have been found in previous theoretical
studies of weak incompressible MHD turbulence (Bhattacharjee \& Ng
1997, Goldreich \& Sridhar~1997, Galtier et~al~2000, Bhattacharjee \&
Ng~2001, Perez \& Boldyrev~2008), as well as strong isotropic MHD
turbulence with cross helicity (Grappin et~al~1983). In this paper, it
is argued that spectral indices~$> 5/3$ are a consequence of cross
helicity in strong anisotropic MHD turbulence.

\begin{figure}[h]
\includegraphics[width=3.5in]{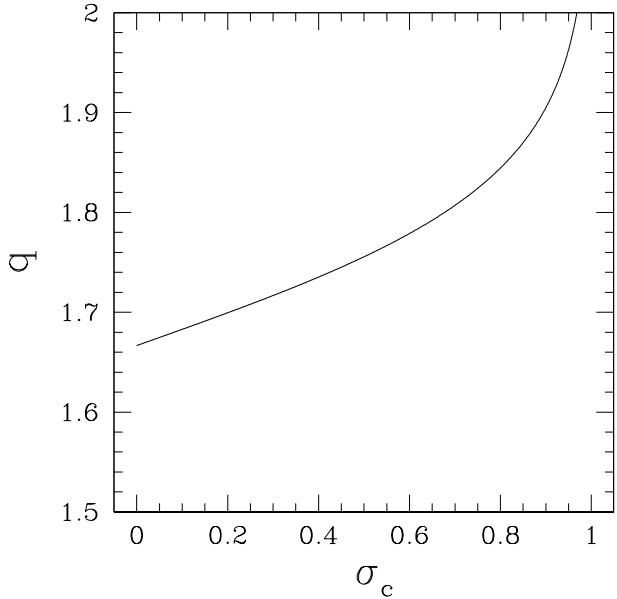}
\caption{\footnotesize Dependence of spectral index of the total
  energy spectrum,~$q$, on the outer-scale fractional cross
  helicity~$\sigma_c$. The ratio of the dissipation wavenumber~$k_d$
  to the perpendicular wavenumber at the outer-scale~$k_f$ in
  equation~(\ref{eq:defc32}) is taken to be~$3780$.
\label{fig:f5}}
\end{figure}

The simplest way to apply this paper to solar wind turbulence is to
model the solar wind fluctuations at some location as steady-state,
forced, homogeneous turbulence with the same average value of
$w_{k_f}^+/w_{k_f}^-$, where $w_{k_f}^+$ and $w_{k_f}^-$ are the rms
amplitudes of the $w^\pm$ fluctuations at the outer scale~$k_f^{-1}$.
Upon setting $w_{k_\perp}^\pm \propto k_\perp ^{-a^\pm}$, one can
write $(w^+_{k_\perp}/w^-_{k_\perp})^2 \propto (k_\perp/k_d)^{-2a^+ +
  2a^-}$, where it is assumed that the spectra are equal at the
dissipation wave number~$k_d$.  Equation (\ref{eq:apam1}) then gives
$(w^+_{k_\perp}/w^-_{k_\perp})^2 \simeq (k_\perp/k_d)^{1-3a^+}$, and
the value of $a^+$ can be obtained from the equation
$(w^+_{k_f}/w^-_{k_f})^2 \simeq (k_f/k_d)^{1-3a^+}$. The total-energy
spectrum, $E(k_\perp) = E^+(k_\perp) + E^-(k_\perp)$, is
approximately~$k_\perp^{-1} (w_{k_\perp}^+)^2$, although it is flatter
than $k_\perp^{-1} (w_{k_\perp}^+)^2$ near the dissipation scale where
the flatter spectrum of the~$w^-$ fluctuations is important. Thus, at
scales much larger than the dissipation scale,
\begin{equation}
E(k_\perp) \propto k_\perp^{-q},
\end{equation} 
where
\begin{equation}
q = \frac{5}{3} +  \frac{2 \log_{10}[(w_{k_f}^+)^2/ (w_{k_f}^-)^2]}{3\log_{10}(k_d/k_f)}.
\label{eq:defc3} 
\end{equation} 
  Upon defining
the outer-scale fractional cross helicity as
\begin{equation}
\sigma_c = \frac{(w_{k_f}^+)^2 - (w_{k_f}^-)^2}{(w_{k_f}^+)^2 + (w_{k_f}^-)^2}
\label{eq:defsigmac} 
\end{equation} 
one can rewrite equation~(\ref{eq:defc3}) as
\begin{equation}
q =  \frac{5}{3} + \frac{2 \log_{10}[(1+ \sigma_c)/ (1- \sigma_c)]}{3\log_{10}(k_d/k_f)}.
\label{eq:defc32} 
\end{equation} 
The spectral index from equation~(\ref{eq:defc32}) is plotted in
Figure~\ref{fig:f5}, assuming that $k_d/k_f = f_d/f_b = 3780$, where
$f_b = (3.5 \mbox{ hours})^{-1}$ is the break frequency at 1~AU
discussed above (Matthaeus \& Goldstein 1986) and $f_d = 0.3 \mbox{
  s}^{-1}$ is the frequency at the dissipation scale. (Smith
et~al~2006) With this choice, $q=1.78$ for $(w_{k_f}^+)^2/ (w_{k_f}^-)^2
= 4$ and $q=1.74$ for $(w_{k_f}^+)^2/ (w_{k_f}^-)^2 = 2$.  When
equation~(\ref{eq:defc32}) is applied to the solar wind, $\sigma_c$
should be interpreted as the cross helicity at the outer
scale~$k_f^{-1}$ averaged over at least a few outer-scale
fluctuations.

Some caution, however, is warranted when applying
equation~(\ref{eq:defc32}) to the solar wind because solar wind
conditions vary with~$r$, while equation~(\ref{eq:defc32}) is based on
results for homogeneous turbulence. The condition under which it is
valid to treat the solar-wind fluctuations at a some~$r$ with an
observed~$\sigma_c$ as homogeneous turbulence with the same $\sigma_c$
is that the cascade time at the outer scale $\tau_{k_f}^\pm$ be much
smaller than the time $t_{\rm adv} = r/U$ for turbulent structures to
be advected a distance~$r$, where $U$ is the solar wind speed.
However, this condition is often either violated or only marginally
satisfied.  This is illustrated by the results of Matthaeus \&
Goldstein (1982) based on four days of Voyager data (in their
Table~1).  For the fluctuations at 1~AU, these authors found an rms
velocity fluctuation of~$\delta v = 25.5$~km/s [which is comparable
  to~$v_A$ at 1~AU], a velocity correlation length (outer scale) of
$L_\perp \sim 2.83\times 10^{11}$~cm, and an average solar-wind speed
of $352$~km/s.  In this particular data set, $\sigma_c = 0.06$, so
that $w_{k_f}^+ \sim w_{k_f}^-$, with $\tau_{k_f}^+ \sim \tau_{k_f}^-
\sim L_\perp /\delta v = 1.11 \times 10^5$~s. On the other hand,
$t_{\rm adv} = \mbox{ (1 AU)}/U = 4.26 \times 10^5$~s, which
marginally satisfies the condition $t_{\rm adv} \gg \tau_{k_f}^+$.
However, if $w_{k_f}^+ /w_{k_f}^-$ were instead~$\sim 3$ at 1~AU, as
in the results of Bavassano et~al~(2000), then $\tau_{k_f}^+$ would
be somewhat larger than~$t_{\rm adv}$. Thus, the inhomogeneity of
the solar wind may influence the effects of cross helicity on the
spectral indices. However, more detailed modeling of inhomogeneous
solar-wind turbulence is beyond the scope of this paper.

\section{Comparison to Other Studies}
\label{sec:comp} 

In this section, the results of this paper are compared to 
two recent studies of strong anisotropic incompressible MHD turbulence
with cross helicity.

\subsection{Lithwick, Goldreich, \& Sridhar (2007) }

The model for the cascade of energy to larger~$k_\parallel$ used in
this paper is based on the results of Lithwick, Goldreich, \&
Sridhar~(2007) (hereafter LGS07). As a result, in both studies, if
$w^+$ and $w^-$ have comparable correlation lengths in the direction
of~$\bm{B}_0$ at the outer scale, then $\overline{
  k}_\parallel^{\,+}\simeq \overline{ k}_\parallel^{\,-}$ at all
smaller scales. The principal difference between this paper and~LGS07
lies in our treatment of the cascade time for the dominant fluctuation
type,~$w^+$. LGS07 argue that if $w_{k_\perp}^+ \gg w_{k_\perp}^-$,
$\chi_{k_\perp}^+ \sim 1$, and $\chi_{k_\perp}^- \ll 1$, then the
shearing applied by $w^-$ wave packets on a $w^+$ wave packet at
perpendicular scale~$k_\perp^{-1}$ is coherent over a time~$(k_\perp
w_{k_\perp}^-)^{-1}$, which greatly exceeds the time $ (\overline{
  k}_\parallel^{\,-} v_A)^{-1}$ required for a $w^+$ and $w^-$ wave
packet at perpendicular scale $k_\perp^{-1}$ to pass through each
other.  In contrast, in this paper, it is argued that the coherence
time for the straining of the $w^+$ wave packet is of order the
``crossing time'' $ (\overline{ k}_\parallel^{\,-} v_A)^{-1}$. As a
result, the results obtained in this paper for the inertial-range
power spectra, degree of anisotropy, cascade time, and energy fluxes
are different from those of~LGS07.

The approach taken in this paper is motivated by the following
argument.  As argued by LGS07 and Maron \& Goldreich~(2001), the $w^-$
fluctuations propagate approximately along the hypothetical magnetic
field lines obtained from the sum of~$\bm{B}_0$ and the magnetic field
of the~$w^+$ fluctuations.  Let us call these hypothetical magnetic
field lines the ``$w^+$ field lines,'' and let us consider a $w^+$
wave packet of perpendicular scale~$k_\perp^{-1}$ and parallel scale
$(\overline{ k}_\parallel)^{-1}$, where $\overline{ k}_\parallel =
\overline{ k}_\parallel^+ \simeq \overline{ k}_\parallel^-$.  Let us
work in a frame of reference that moves at speed~$v_A$ in the~$-z$
direction along with the~$w^+$ fluctuations. Let us also take an
initial snapshot of the turbulence at $t=0$ and trace out all of the
``$w^+$ field lines'' that pass through our wave packet. The volume
filled by these $w^+$~field lines is the ``source region'' from which
the $w^-$ wavepackets encountered by our $w^+$ wave packet
originate. If we wait one crossing time $(\overline{ k}_\parallel
v_A)^{-1}$ and take a new snapshot of the turbulence, then at any
given location the $w^+$ fluctuations will not have changed very much,
since $w_{k_\perp}^- \ll w_{k_\perp}^+$. However, if we trace out the
new $w^+$ field lines passing through our wave packet, the volume that
is filled by these new $w^+$ field lines will differ substantially
from the initial source region at distances~$\ga (\overline{
  k}_\parallel)^{-1} $ from our wave packet due to the rapid
divergence of neighboring field lines in MHD turbulence.  In other
words, small local changes in $w^+$ lead to large changes in the
connectivity of the $w^+$ field lines.

To see this, let the $w^+$ field line that passes through some
point~$P$ in our wave packet at~$t=0$ be called ``field line~$A$.''
Let the $w^+$ field line that passes through point~$P$ at $t=
(\overline{ k}_\parallel v_A)^{-1}$ be called ``field line~$B$.''  As
before, let us work in a frame of reference that moves at speed~$v_A$
in the~$-z$ direction. Field lines~$A$ and~$B$ are fixed curves, since
they are traced out within two snapshots of the turbulence.  If we
follow field line~$B$ for a distance~$\ll (\overline{
  k}_\parallel)^{-1}$, it will separate from field line~$A$ by some
small distance~$x$ that is~$\ll k_\perp^{-1}$.  If we continue to
follow field line~$B$, its separation from field line~$A$ is analogous
to the separation of two neighboring field lines within a single
snapshot of the turbulence. As shown by Narayan \& Medvedev~(2001),
Chandran \& Maron~(2004), and Maron, Chandran, \& Blackman~(2004), if
a pair of field lines is separated by a distance~$x \ll k_\perp^{-1}$
at one location, then the distance the field-line pair must be
followed before it separates by a distance~$k_\perp^{-1}$ is a few
times the parallel size of an eddy of perpendicular
size~$k_\perp^{-1}$ - i.e., a few times~$(\overline{
  k}_\parallel)^{-1}$. It turns out that the particular value of $x$
has little effect unless one considers the (irrelevant) case in which
$x/d \sim e^{-N},$ where $N$ is large and $d$ is the perpendicular
dissipation scale. (Chandran \& Maron 2004) This is because within the
inertial range the amount of magnetic shear increases towards small
scales; therefore, if $x$ is made very small, then the distance one
has to follow the field-line pair in order for~$x$ to double becomes
very small. Thus, as a result of the rapid divergence of neighboring
field lines in MHD turbulence, the ``source region'' of our $w^+$ wave
packet at $t=(\overline{ k}_\parallel v_A)^{-1}$ differs substantially
from the source region at $t=0$ at distances $\ga (\overline{
  k}_\parallel)^{-1}$ from our wave packet. Because the $w^-$
fluctuations vary rapidly in the direction perpendicular to the
magnetic field, the $w^-$ wave packets encountered by our $w^+$ wave
packet will decorrelate on a time scale of order the crossing time
$(\overline{ k}_\parallel v_A)^{-1}$ due to the time evolution of the
source region.

It should noted that there are two unexplained aspects of LGS07's
model, as pointed out by Beresnyak \& Lazarian~(2007).  The first
concerns the nature of the transition from the weak-turbulence
regime~($\chi_{k_\perp}^+ \ll 1$ and $\chi_{k_\perp}^- \ll 1$) to the
strong-turbulence regime~($\chi_{k_\perp}^+\sim 1$ and
$\chi_{k_\perp}^- \ll 1$).  In LGS07's analysis, as one passes from
the weak regime to the strong regime, the coherence time for the
straining of the $w^+$ wave packets by~$w^-$ wave packets increases by
a factor of $(\chi_{k_\perp}^-)^{-1}$ and the energy cascade
time~$\tau_{k_\perp}^+$ decreases by a factor of~$\chi_{k_\perp}^-$.
It is not clear why these large changes should occur across the
transition scale.  The second issue is that LGS07 find that
$E^+(k_\perp) \propto k_\perp^{-5/3}$ and $E^-(k_\perp) \propto
k_\perp^{-5/3}$ regardless of the fractional cross helicity.  Since
the ratio $E^+(k_\perp)/E^-(k_\perp)$ is independent of wavenumber, it
is not clear how pinning could occur in their model, or how the
spectra would behave near the dissipation scale.

\subsection{Beresnyak \& Lazarian (2007)}

Beresnyak \& Lazarian (2007) (hereafter BL07) have published an online
article on strong MHD turbulence with cross helicity. The following
discussion refers to the version of their article that is available
electronically as of the writing of this paper. The work of BL07 is
similar to this paper in that both studies take the dominant
fluctuation type, $w^+$, to undergo a weak cascade. Also,
equations~(\ref{eq:defapam}) through (\ref{eq:olkp1}) of this paper
are equivalent to their equation~(5), except for the fact that they
take the parallel correlation lengths of $w^+$ and $w^-$ to differ by
a constant multiplicative factor when a power-law solution for the
spectra is assumed. (The possibility of more general solutions is
claimed by BL07.) On the other hand, there are a number of significant
differences between this paper and Beresnyak \& Lazarian's (2007)
work. They argue that for the $w^+$ fluctuations, the dominant
nonlinear interactions are between fluctuations with comparable
parallel correlation lengths and different perpendicular correlation
lengths, whereas for the $w^-$ fluctuations the dominant interactions
are between fluctuations with comparable perpendicular scales.
Here, it is argued that for both $w^+$ and $w^-$ the
dominant interactions are between fluctuations with similar
perpendicular scales.  When $w^+_{k_\perp} \gg w^-
_{k_\perp}$ and $\chi_{k_\perp}^+ \sim 1$, their Figure~1 suggests
that the parallel correlation length of~$w^+$ fluctuations can be less
than the parallel correlation length of~$w^-$ fluctuations.  It is
argued in section~\ref{sec:unequal} of this paper that this can not be
the case, because the $w^+$ fluctuations will imprint their parallel
correlation length onto the~$w^-$ fluctuations.  They argue that the
scalings given by equations~(\ref{eq:defapam}) and (\ref{eq:olkp1})
[equivalently, their equation~(5)] can not apply if the parallel
correlation lengths of the $w^+$ and $w^-$ fluctuations are equal,
arguing that this would require $\epsilon^+ = \epsilon^-$, whereas in
this paper the ratio~$\epsilon^+/\epsilon^-$ depends upon the slopes
of the power spectra, as in weak turbulence.  They argue that if the
$w^+$ and $w^-$ fluctuations are driven with the same parallel
correlation length at the outer scale, there will be a non-power law
part of the solution at large scales that will transition at smaller
scales to a power law solution with $w_{k_\perp}^+$ and
$w^-_{k_\perp}$ both $\propto k_\perp^{-1/3}$ and $\overline{
  k}_\parallel ^{\,\pm}\propto k_\perp^{2/3}$. In contrast, in this
paper a power-law solution starting at the outer scale is obtained
with different scalings for $w^+_{k_\perp}$ and $w^-_{k_\perp}$, and
with $\overline{ k}_\parallel^{\pm}$ growing more slowly
than~$k_\perp^{2/3}$.

\section{Conclusion}

This paper proposes a new phenomenology for strong, anisotropic,
incompressible MHD turbulence with cross helicity and introduces a
nonlinear advection-diffusion equation [equation~(\ref{eq:FPpm})] to
describe the time evolution of the anisotropic power spectra of the
$w^+$ and $w^-$ fluctuations.  It is found that in steady state the
one-dimensional power spectra of the energetically dominant $w^+$
fluctuations, $E^+(k_\perp)$, is steeper than $k_\perp^{-5/3}$, and
that $E^+(k_\perp)$ becomes increasingly steep as the fractional cross
helicity~$\sigma_c$ increases.  Increasing $\sigma_c$ also increases
the energy cascade time of the $w^+$ fluctuations, reduces the
turbulent heating power for a fixed fluctuation energy, and increases
the anisotropy of the fluctuations at small scales.

Although most of the discussion has focused on forced, steady-state
turbulence, the results of this paper can also be applied to decaying
turbulence.  For example, equations~(\ref{eq:tau1}) and
(\ref{eq:tau2}) can be used to estimate the time scale for turbulence
to decay.  The resulting prediction is that if the fluctuations are
initially excited with $w_{k_f}^+ \gg w_{k_f}^-$ and with comparable
parallel correlation lengths at the outer scale, then the turbulence
will decay into a state in which~$w^-$ fluctuations are absent, as in
the earlier work of Dobrowolny, Mangeney, \& Veltri (1980), Grappin
et~al~(1983), and Lithwick \& Goldreich (2003).  This ``maximally
aligned'' state will then be free from nonlinear interactions, and
will persist for long times until it damps via linear dissipation.

\acknowledgements I thank Chuck Smith, Sebastien Galtier, Bernie
Vasquez, Yoram Lithwick, Alex Lazarian, and Andrey Beresnyak for
helpful comments and suggestions.  This work was supported in part by
the NSF/DOE Partnership in Basic Plasma Science and Engineering under
grant number AST-0613622, by NASA under grant numbers NNX07AP65G and
NNH06ZDA001N-SHP06-0071, and by DOE under grant number
DE-FG02-07-ER46372.

\references

Antonucci, E., Dodero, M. A., \& Giordano, S. 2000, Sol. Phys.,
197, 115 
                     
Barnes, A. 1966, Phys. Fluids, 9, 1483

Bavassano, B., Pietropaolo, E., \& Bruno, R. 2000, J. Geophys. Res.,
105, 15959

Beresnyak, A., \& Lazarian, A. 2006, ApJ, 640, L175

Beresnyak, A., \& Lazarian, A. 2007, arXiv:0709.0554v1

Bhattacharjee, A. \& Ng, C. S. 2001, ApJ 548, 318 

Bieber, J., Matthaeus, W., Smith, C., Wanner, W., Kallenrode, M., \&
Wibberenz, G. 1994, ApJ, 420, 294

Biskamp, D., Schwarz, E., \& Drake, J. F. 1996, Phys. Rev. Lett., 76, 1264

Biskamp, D., Schwarz, E., Zeiler, A., Celani, A., \& Drake, J. F. 1999, Phys. Plasmas, 6, 751

Boldyrev, S., Nordlund, A., \& Padoan, P. 2002, Phys. Rev. Lett., 89,
031102

Boldyrev, S. 2005, ApJ, 626, L37

Boldyrev, S. 2006, Phys. Rev. Lett., 96 115002

Brodin, G., Stenflo, L., \& Shukla, P. K. 2006, Solar Phys., 236,
285

Bruno, R., \& Carbone, V. 2005, Living Rev. Solar Phys., 2, 4
[Online article: cited Nov. 14, 2007, http://www.livingreviews.org/lrsp-2005-4]

Chandran, B. 2004, Sp. Sci. Rev., 292, 17

Chandran, B. 2005, Phys. Rev. Lett., 95, 265004

Cho, J., \& Vishniac, E. 2000, ApJ, 539, 273

Cho, J., Lazarian, A., \& Vishniac, E. 2002, ApJ, 564, 291

Cho, J., \& Lazarian, A. 2002, Phys. Rev. Lett., 88, 245001

Cho, J., \& Lazarian, A. 2003, MNRAS, 345, 325

Cho, J., \& Lazarian, A. 2004, ApJL, 615, L41

Cranmer, S. R. \& van Ballegooijen, A. A. 2005, ApJS, 156, 265

Cranmer, S. R. \& van Ballegooijen, A. A. 2007, ApJS, 171, 520

Dmitruk, P., Milano, L. J., \&
Matthaeus, W. H. 2001, ApJ, 548, 482

Dmitruk, P., Matthaeus, W. H., Milano, L. J., Oughton, S., Zank, G. P.,
\& Mullan, D. J. 2002, ApJ, 575, 571

Dobrowolny, M., Mangeney, A., Veltri, P. L. 1980, Phys. Rev. Lett., 35, 144

Elmegreen, B. G.,  \& Scalo, J., ARAA, 42, 211

Galtier, S., Nazarenko, S. V., Newell, A. C., \& Pouquet,
A. 2000, J. Plasma Phys., 63, 447

Galtier, S., \& Chandran, B. 2006, Phys. Plasmas, 3, 114505

Galtier, S., \& Buchlin, E. 2007, ApJ, 656, 560

Goldreich, P., \& Sridhar, S. 1995, ApJ, 438, 763

Goldreich, P., \& Sridhar, S. 1997, ApJ, 485, 680

Goldstein, B. E., Smith, E. J., Balogh, A., Horbury, T. S., Goldstein, M. L.,
\& Roberts, D. A. 1995, Geophys. Res. Lett., 22, 3393

Grappin, R., Pouquet, A., \& L\'eorat, J. 1983, A\&A, 126, 51

Haugen, N. E. L., Brandenburg, A., \& Dobler, W., Phys. Rev. E. 2004,
70, 016308

Higdon, J. 1984, ApJ, 285, 109

Howes, G. G., Cowley, S. C., Dorland, W., Hammett, G. W., Quataert, E.,
\& Schekochihin, A. A. 2006, ApJ,  651  590-614

Howes, G. G., Cowley, S. C., Dorland, W., Hammett, G. W., Quataert, E.,
\& Schekochihin, A. A. 2007a, arXiv:0707.3147

Howes, G. G., Dorland, W., Cowley, S. C.,  Hammett, G. W., Quataert, E.,
Schekochihin, A. A., \& Tatsuno, T. 2007b, arXiv:0711.4355

Iroshnikov, P. 1963, Astron. Zh. 40, 742

Klein, L. W., Matthaeus, W. H., Roberts, D. A., \& Goldstein, M. L. 1992,
{\em Solar Wind Seven; Proceedings of the 3rd COSPAR Colloquium, Goslar, Germany,} pp. 197-200.

Kohl, J. L. et al 1998, ApJL, 501, 127

Kolmogorov, A. N. 1941, Dokl. Akad. Nauk SSSR, 30, 299

Kraichnan, R.  H. 1965, Phys. Fluids 8, 1385

Kuznetsov, E. A. 2001, J. Exp. Theor. Phys., 93, 1052

Leith, C. E., \& Kraichnan, R. H. 1972, J. Atmos. Sci., 29, 1041

Lithwick, Y., \& Goldreich, P. 2001, ApJ, 562, 279

Lithwick, Y., \& Goldreich, P. 2003, ApJ, 582, 1220

Lithwick, Y., Goldreich, P., \& Sridhar, S. 2007, ApJ, 655, 269 (LGS07)

Luo, Q., \& Melrose, D. 2006, MNRAS, 368, 1151

Maron, J., Chandran, B., \& Blackman, E. 2004, Phys. Rev. Lett., 92,
045001

Maron, J., \& Goldreich, P. 2001, ApJ, 554, 1175

Marsch, E., \& Tu C.-Y. 1996, J. Geophys. Res., 101, 11149

Mason, J., Cattaneo, F., \& Boldyrev, S. 2006, Phys. Rev. Lett., 97, 255002

Matthaeus, W. H., \& Goldstein, M. L. 1982, J. Geophys. Res., 87, 6011

Matthaeus, W. H., \& Goldstein, M. L. 1986, Phys. Rev. Lett., 57, 495

Matthaeus, W. H., \& Montgomery, D. 1980,  New York Acad. Sci., 357, 203

Matthaeus, W. H., Zank, G. P., Leamon, R. J.,
Smith, C. W., Mullan D. J.,  \& Oughton, S. 1999, Sp. Sci. Rev., 87, 269

Matthaeus, W. H., Mullan D. J., Dmitruk, P., Milano, L., \& Oughton, S. 2002,
Non. Proc. Geophys., 9, 1

Matthaeus, W. H., Dmitruk, P., Smith, D., Ghosh, S., \& Oughton, S. 2003,
Geophys. Res. Lett., 30, 4-1

Matthaeus, W. H., Pouquet, A., Mininni, P. D., Dmitruk, P.,
\& Breech, B. 2007, arXiv:0708.0801

Mininni, P., \& Pouquet, A. 2007, Phys. Rev. Lett., 99, 254502 

Moffatt, H. K. 1978, {\em Magnetic field generation in electrically conducting fluids}  (Cambridge, England: Cambridge University Press, 1978)

Montgomery, D., \& Matthaeus, W. 1995, ApJ, 447, 706

Montgomery, D., \& Turner, L. 1981, Phys. Fluids, 24, 825

M\"uller, W. C., \& Biskamp, D. 2000, Phys. Rev. Lett., 84, 475

M\"uller, W. C., Biskamp, D., \& Grappin, R. 2003, Phys. Rev. E, 67, 066302

M\"uller, W. C., \& Grappin, R. 2005, Phys. Rev. Lett., 95, 114502

Ng, C. S., \& Bhattacharjee, A. 1996, ApJ, 465, 845

Ng, C. S., \& Bhattacharjee, A. 1997, Phys. Plasmas, 4, 605

Oughton, S., Matthaeus, W. H., \& Ghosh, S. 1995, LNP, Vol.~462: {\em
  Small-Scale Structures in Three-Dimensional Hydrodynamic and
  Magnetohydrodynamic Turbulence}, p. 273

Oughton, S., Dmitruk, P., \& Matthaeus, W. H. 2006, Phys. Plasmas,
13, 2306

Padoan, P., Jimenez, R., Nordlund, A., \& Boldyrev, S. 2004, Phys.
Rev. Lett., 92, 191102

Perez, J. C., \& Boldyrev, S. 2008, ApJL, 672, L61

Podesta, J. J., Roberts, D. A., \& Goldstein, M. L. 2007, ApJ, 664, 543

Pouquet, A., Sulem, P. L., \& Meneguzzi, M. 1988, Phys. Fluids, 31, 2635

Roberts, D. A., Goldstein, M. L., Klein, L. W., \& Matthaeus, W. H. 1987,
J. Geophys. Res., 92, 12023

Schekochihin, A. A., Cowley, S. C., Dorland, W., Hammett, G. W., Howes, G. G., Quataert, E., \&  Tatsuno, T 2007b, arXiv:0704.0044

Shebalin, J. , Matthaeus, W., \& Montgomery, D. 1983, J. Plasma Phys., 
29, 525

Shukla, P. K., Brodin, G., \& Stenflo, L. 2006, Phys. Lett. A,
353, 73

Skilling, J., McIvor, I., \& Holmes, J. 1974, MNRAS, 167, 87P

Smith, C. W. 2003, {\em Solar Wind Ten: Proceedings of the Tenth
  International Solar Wind Conference}, AIP Conference Proceedings,
679, 413

Smith, C. W., Hamilton, K., Vasquez, B., \& Leamon, R. 2006, ApJL, 645, 85

Stone, J. M., Ostriker, E. C., \&  Gammie, C. F. 1998, ApJL, 508, 99

Taylor, G. I. 1938, Proc. R. Soc. London A, 164, 476

Velli, M. 1993, A\&A, 270, 304

Verdini, A., \& Velli, M. 2007, ApJ, 662, 669

Zhou, Y., \& Matthaeus, W. H. 1990, J. Geophys. Res., 95, 10291

\end{document}